# Co-Enrichment of Proteins in Extracellular Vesicles

Molly L. Shen[1-4], Andreas Wallucks[1,2], Rosalie Martel[1,2,†], Zijie Jin[1,2,†], Lucile Alexandre[1,2], Philippe DeCorwin-Martin[1,2], Lorenna Oliveira Fernandes de Araujo[1,2], Andy Ng[1,2], Peter M. Siegel[3,4], and David Juncker[1,2,*]

[1]Biomedical Engineering Department, McGill University, Montréal, QC, Canada

[2]Victor Phillip Dahdaleh Institute of Genomic Medicine, McGill University, Montréal, QC, Canada

[3]Goodman Cancer Institute, McGill University, Montréal, QC, Canada

[4]Department of Medicine, McGill University, Montréal, QC, Canada

[†] R.M. and Z.J. contributed equally.

***Abstract:***

Extracellular vesicles (EVs) are cell-derived secretions that mediate tissue homeostasis and intercellular communication through their diverse cargos, such as proteins. Distinct EV biogenesis pathways suggest specific association and co-enrichment of proteins sharing a biogenesis pathway, and non-association and co-depletion of proteins segregated into distinct pathways. Yet these associations elude conventional protein expression or co-expression measurements. Here, we propose and define pairwise protein co-enrichment (CoEn) to quantify whether a given protein is co-enriched or co-depleted with another protein relative to its overall expression. We measure CoEn, and differential CoEn (dCoEn) between a stimulus and a reference condition, of up to 240 protein pairs in EVs using antibody microarrays. We validate CoEn by modulating well-known EV biogenesis pathways, and find that dCoEn quantifies expected changes between perturbed and reference conditions while uncovering new ones; CoEn and dCoEn in three model cell lines and parental and organotropic breast cancer progeny cell lines reveals both preserved and variable CoEn that may warrant further studies. Collectively, our result suggest that CoEn reflects and illuminates cell physiology and EV biogenies, is readily measurable, and could further serve as quality control in EV biomanufacturing as well as underpin new EV biomarkers.

## 1. Introduction:

Extracellular vesicles (EVs) are membrane-derived, cargo-carrying vesicles secreted by all cell types that are found in various body fluids and have attracted large amounts of interest after it was shown numerous times that they serve as a bona fide communication system in the body.[1–5] EVs are transported via the blood circulatory system and their extensive involvement in health and disease makes them an attractive source of biomarkers for liquid biopsy,[6–8] as well as vehicles for drug delivery.[9,10] EVs are broadly categorized into two main types based on their biogenesis: exosomes and ectosomes (also known as microvesicles). Exosomes are formed within the endosomal system and released upon fusion of multivesicular bodies (MVBs) with the plasma membrane, whereas ectosomes bud directly from the plasma membrane. Beyond the EV membrane of origin, distinct cargo-sorting mechanisms (or EV biogenesis pathways) contribute to the widely observed heterogeneity in EV molecular cargoes. The maybe best-characterized cargo pathways are the ESCRT-dependent pathway and the Syntenin-ALIX pathway.[11] Both require the highly coordinated assembly and disassembly of protein complexes that orchestrate membrane budding and protein cargo loading via specific motifs (e.g., ubiquitination) or protein interactions (e.g., Syndecan binding). Other cargo pathways exist where vesicle budding is based on biophysical mechanisms, notably the ceramide pathways or ectosome budding via cytoskeletal remodeling. While the exact mechanisms for these remain less well understood, it is often hypothesized that they result in less-specific cargo loading, relying more on bulk capture or association to membrane microdomains.

A long-standing challenge in the field is to precisely map specific EV protein cargoes to their respective biogenesis pathways, to quantify each pathway's relative contribution to the overall EV secretion by cells, and to characterize their differential regulation under various physiological and pathological conditions. Currently, there are neither universal nor definitive markers of EV biogenesis pathways. Nevertheless, observations established that various protein cargos are enriched in particular EV types (e.g., in exosome or in ectosome). For example CD63 is known to be predominantly enriched in exosomes originating from endosomes, whereas CD147 is enriched in ectosomes derived from the plasma membrane.[12] Furthermore, pairs of proteins are expected to be co-enriched or co-depleted inside EVs due to either direct interactions (e.g., ALIX with Syntenin-1 based on their well-known binding) or their abundance at specific sites of EV biogenesis (e.g., integrins clustering in distinct membrane regions without direct binding). Following this logic, proteins that are enriched in the same type of EVs because of partially or fully shared biogenesis pathways are expected to be co-enriched in individual EVs, and less frequently found individually without their co-enrichment partner. Importantly, pairs of proteins can be co-enriched without being universal pathway markers, i.e., proteins sharing a cargo sorting pathway will be often found together even if they are both individually rare in the total population. Conversely, proteins that shared distinct pathways could be negatively co-enriched, or co-depleted.

Co-enrichment metrics are widely used across disciplines to quantify the relative overrepresentation and underrepresentation of features within subsets of data. In genomics, co-enrichment measures the overrepresentation of features or feature combinations and helps define distinct subpopulations in clustering analysis - for example, identifying marker genes enriched in single-cell RNA-seq clusters to annotate cell types.[13] In medicine, the odds ratio measures how much more likely an outcome is in the presence of a particular condition or treatment plan.[14] Outside of biology, recommendation systems identify products that are "frequently purchased together" and inform personalized online shopping experiences. We reasoned that a quantitative, EV-specific co-enrichment metric could similarly offer new insights into how tightly EV cargo packaging is regulated, reveal how proteins are dynamically re-routed between EV biogenesis pathways in response to cellular stimuli or stress, and support the development of classifiers for EV subpopulations and EV-based biomarkers.

Here, we define a normalized metric for quantifying the pairwise co-enrichment of a protein A with a protein B within EVs which we term CoEn. CoEn is formally defined as the logarithm of the co-expression (CoEx) of a protein A in $B^+$ EVs, normalized by the global expression (Ex) of the protein A in the entire EV population. We demonstrate an experimental workflow to generate CoEn measurements in bulk EV analysis using immunofluorescence on an antibody microarray by (i) measuring the CoEx of EVs on antibody capture spots, (ii) measuring the Ex in the entire EV sample using a pan-EV capture surface, and (iii) deriving CoEn by normalizing CoEx with Ex. To confirm CoEn's validity and utility, we measured changes in CoEn upon perturbating two well-known biological mechanisms and pathways: (i) the selective re-routing of Syntenin-1 upon blocking binding to its partner CD63 and (ii) systemic protein re-routing in EVs upon inhibition of lysosomal acidification and protein degradation by bafilomycin A1. We compared the changes in Ex, CoEx, and CoEn under these conditions and found that CoEn reflected the expected cellular response and further uncovered previously unknown responses; yet the changes were detectable by neither Ex nor CoEx measurements. We then profiled CoEn of 240 protein pairs in EVs in three distinct model cell lines (293T, HT29 and MDA-MB-231). Cell line and cancer-specific CoEn were analyzed by hierarchical clustering and principal component analysis (PCA), identifying conserved sets of proteins that mutually co-enrich across all three cell lines as well as sets that are co-depleted. Finally, measured the CoEn of EVs of organotropic breast cancer cell lines (i.e., bone-, lung-, liver-, and brain-tropic), and found CoEn features unique to each organotropism, hinting at CoEn's potential for biomarker discovery. Collectively, the results establish CoEn as an analytical tool and a generator of hypothesis.

## 2. **Results:**

### *2.1 Conceptual and Experimental Framework for Protein Co-enrichment in EVs*

We define the pairwise CoEn of a protein A in a $B^+$ subpopulation of EVs ($\mathrm{CoEn}_{B^+}^{A}$) as

$$\mathrm{CoEn}_{B^+}^{A} = \log_{10}\left(\frac{\mathrm{CoEx}_{B^+}^{A}}{\mathrm{Ex}^{A}}\right). \tag{1}$$

with $\mathrm{CoEx}_{B^+}^{A}$ the co-expression of protein A in $B^+$ EVs and $\mathrm{Ex}^{A}$ the global expression of protein A in the entire EV population. CoEn values can be positive, negative (for co-depletion) or zero for no co-enrichment, Figure 1A. To help visualize and concretize CoEn, we show a hypothetical, oversimplified scenario of pairwise CoEn combinations for 4 proteins, namely protein A and B that are preferentially associated with pathway 1, protein C which follows pathway 2, and protein D which associated with both pathways, Figure 1B.

We define $\mathrm{CoEx}_{B^+}^{A}$ is the average number of protein A on EVs that are also positive for protein B

$$\mathrm{CoEx}_{B^+}^{A} = \frac{N_{B^+}^{A}}{n_{B^+}}, \tag{2}$$

where $N_{B^+}^{A}$ is the copy number of protein A on $B^+$ EVs and $n_{B^+}$ is the number of $B^+$ EVs in the sample. Conversely, $\mathrm{Ex}^{A}$ is the average number of protein A in all EVs

$$\mathrm{Ex}^{A} = \frac{N^{A}}{n}, \tag{3}$$

where $N^{A}$ is the copy number of protein A and n is the sample size. If loading of protein A was a purely random process, $\mathrm{CoEx}_{B^+}^{A}$ and $\mathrm{Ex}^{A}$ would be equal and hence CoEn = 0. We therefore propose CoEn ≠ 0 as indicator of non-random association (co-enrichment or co-depletion) of the proteins A and B: a positive CoEn (CoEn >0) indicate that a protein A is enriched in the $B^+$ subpopulation, whereas a negative CoEn (CoEn <0) indicate that a protein A is depleted in $B^+$ EVs, as illustrated in Figure 1.

Throughout this manuscript, we measure both $\mathrm{CoEx}_{\mathrm{B}^+}^{\mathrm{A}}$ and $\mathrm{Ex}^{\mathrm{A}}$ by immunofluorescence on custom antibody microarrays (Figure 2), and we describe how to derive them using a conventional ELISA in the Supplemental Method. To measure $\mathrm{CoEx}_{\mathrm{B}^+}^{\mathrm{A}}$, we selectively capture $B^+$ EVs on microarray capture spots via an anti-B antibody and subsequently measure the detection signal $\mathrm{I}_{\mathrm{B}^+}^{\mathrm{A}}$ of fluorescent anti-A antibodies on the captured $B^+$ EVs. The population size of captured $B^+$ EVs is quantified by the intensity $\mathrm{I}_{\mathrm{B}^+}^{\Omega}$ of a pan-EV labeling dye $\Omega$. The ratio of the two measurements is proportional to $\mathrm{CoEx}_{\mathrm{B}^+}^{\mathrm{A}}$ with a proportionality factor $\alpha$ that is given by the intensity of the fluorescence channels:

$$\mathrm{CoEx}_{\mathrm{B}^+}^{\mathrm{A}} = \alpha \frac{\mathrm{I}_{\mathrm{B}^+}^{\mathrm{A}}}{\mathrm{I}_{\mathrm{B}^+}^{\Omega}}. \tag{4}$$

$\mathrm{Ex}^{\mathrm{A}}$ is accessible in the same antibody microarray setup by introducing a pan-EV capture surface, see below. After EV pulldown, we measure both the detection signal $\mathrm{I}^{\mathrm{A}}$ of anti-A antibodies on all captured EVs and we quantify the capture population size via the intensity $\mathrm{I}^{\Omega}$ of the pan-EV labeling dye. By reusing the same fluorescence channels as for $\mathrm{CoEx}_{\mathrm{B}^+}^{\mathrm{A}}$, the ratio of these two measurements is proportional to $\mathrm{Ex}^{\mathrm{A}}$ with the same proportionality factor $\alpha$:

$$\mathrm{Ex}^{\mathrm{A}} = \alpha \frac{\mathrm{I}^{\mathrm{A}}}{\mathrm{I}^{\Omega}}. \tag{5}$$

Finally, we thus obtain

$$\mathrm{CoEn}_{\mathrm{B}^+}^{\mathrm{A}} = \log_{10}\left(\frac{\mathrm{CoEx}_{\mathrm{B}^+}^{\mathrm{A}}}{\mathrm{Ex}^{\mathrm{A}}}\right) = \log_{10}\left(\frac{\mathrm{I}_{\mathrm{B}^+}^{\mathrm{A}}/\mathrm{I}_{\mathrm{B}^+}^{\Omega}}{\mathrm{I}^{\mathrm{A}}/\mathrm{I}^{\Omega}}\right), \tag{6}$$

where the proportionality factors cancel. We note that as a ratio of two expression levels, CoEn is unitless and can vary independently from the Ex of the protein. Likewise, because the same detection antibody is used for both Ex and CoEx, CoEn is not susceptible to variations in fluorescent labeling efficiency as long as the signals reaches minimal detection thresholds. As such, CoEn is a normalized metric readily comparable across experiments.

## 2.2 Differential CoEn

CoEn furthermore lends itself to analyze changes in protein co-enrichment between EVs collected from (i) cells subjected to a stimulus and from cells under reference conditions, (ii) of EVs from two different cell lines, (iii) or simply of EVs from the same cells taken at two different time points. By subtracting the CoEn measurements of two conditions, it is possible to obtain what we call differential CoEn (dCoEn) as

$$\mathrm{dCoEn}^{\mathrm{A}}_{\mathrm{B+}} = {\mathrm{CoEn}^{\mathrm{A}}_{\mathrm{B+}}}_{\mathrm{stimulus}} - {\mathrm{CoEn}^{\mathrm{A}}_{\mathrm{B+}}}_{\mathrm{reference}} = \log_{10}\left[\left(\frac{\mathrm{CoEx}^{\mathrm{A}}_{\mathrm{B+}}}{\mathrm{Ex}^{\mathrm{A}}}\right)_{\mathrm{stimulus}} \Big/ \left(\frac{\mathrm{CoEx}^{\mathrm{A}}_{\mathrm{B+}}}{\mathrm{Ex}^{\mathrm{A}}}\right)_{\mathrm{reference}}\right]. \quad (7)$$

CoEn being defined as a $\log_{10}$ of the measurements, the subtraction of CoEn values is proportional to the ratio of $\mathrm{CoEx}/\mathrm{Ex}$ of the reference and stimuli condition. dCoEn could be used to measure differences and relative changes between two conditions and changes over time.

## ***2.3 Pan-EV Labeling Dyes***

The calculation of CoEn requires normalization on EV numbers, and hence the accuracy of Ex, CoEx and CoEn values rely on the universality of the pan-EV labels used during the measurements. Fluorescent dyes, such as lipophilic dyes and membrane-permeable organic dyes have been commonly used to visualize and measure EV abundance in bulk and count particles in single EV assays.[15,16] In our study, membrane-permeable organic fluorescent dye Cell Tracker Red (CTR) and the amine-reactive dye 5-(and-6)-Carboxyfluorescein Diacetate Succinimidyl Ester (CFDA-SE) were chosen as candidate universal detection as they are commonly used for non-specific EV labeling, and as they are readily implementable to our assay workflow. N-hydroxysuccinimido biotin (NHS-Biotin)-based EV biotinylation, and subsequent detection via fluorescent streptavidin, were used for cross-validation and benchmarking of CFDA-SE and CTR. We recently introduced label-free size photometry and fluorescence imaging (SPFI) of single EVs, and observed that both CTR and biotin labelled single EVs with high yield–and proportionally to volume and surface area.[17] We confirm the universal labelling property of CTR and of biotinylation for multiple cell lines using SPFI for single EV measurements (Figure 3A and SI Figure 1A) and using microarrays for ensemble EV measurements (SI Figure 1B and 1C). Briefly, purified EVs from breast cancer cell line MDA-MB-231 and colorectal cancer cell line HT29 were dyed with the CTR, immobilized on a coverslip in an unspecific manner, and measured by SPFI. Over 98% of EVs were positive for CTR in both cell lines (Figure 3A and SI Figure 1A), and the fluorescence signal proportional to the size, further supporting proportional CTR labelling of our protocol. We also cross-validated CTR labeling with EV biotinylation using antibody microarray measurements, (SI Figure 1B). The CTR and streptavidin detection signals of 10 antibody-defined EV subpopulations derived from normal kidney cell line 293T were quantified and we observed that CTR and streptavidin labelling followed a similar trend, even for the low abundant subpopulations, affirming CTR's universality. While CFDA-SE was also experimentally validated and is compatible in our workflow (SI Figure 1C), CTR was preferred for its compatibility with our microarray scanner setup.

## *2.4 Pan-EV Capture Probes*

Phosphatidylserine (PS) has been previously used for pan-EV capture,[18,19] but there are widespread concerns that PS is not present on the outer leaflet of all EVs, but only on EVs of cancer cells and apoptotic EVs.[23] Here, we test and validate the use of PS as a pan-EV capture probe by analyzing PS expression (on the outer leaflet) of EVs with FITC-conjugated Annexin V known to bind PS at single EV resolution using SPFI. EVs from four cell lines, including immortalized kidney 293T, breast cancer MDA-MB-231, primary lung fibroblast IMR-90 and colorectal cancer HT29 were physiosorbed on a coverslip and stained with FITC-Annexin V, Figure 3B and 3C, SI Figure 1D. Interestingly, we observed similar PS distributions between the different cell lines (Figure 3B and 3C and SI Figure 1D), where the expression levels of $PS^+$ EVs uniformly scaled with EV size. The global PS positivity furthermore ranged between 62% to 71% among the four cell lines (Figure 3D), which represents a lower bound for PS positivity because we cannot exclude that non-EV particles in the sample were also adsorbed on the coverslip. To further ensure that PS is abundant in various EV subpopulations of interest, we co-detected PS on EVs positive for CD63, LAMP1, CD9 and ITG-β1, each with distinct cellular localizations (i.e., endosomal, lysosomal or membrane-residing) and associated with varying EV sizes, in 293T and MDA-MB-231 EVs (Figure 3E and 3F, SI Figure 2). Consistent with the high global PS abundance, we found well over 74% PS positivity in $CD63^+$, $LAMP1^+$, $CD9^+$ and ITG-$\beta1^+$ EVs in both 293T and MDA-MB-231. Overall, these results support that PS is abundantly expressed in the outer leaflet of EVs of both cancer and non-cancer cell lines, regardless of EV size or protein marker expression.

Next, we incorporated a PS-capture probe in the microarray platform to validate its suitability for CoEn derivations. We chose T-cell immunoglobulin and mucin domain containing 4 (TIM4), known for its high PS binding affinity and its availability as an Fc chimera protein, for pan-EV capture on our microarray platform.[19] Briefly, 293T EVs were biotinylated and dyed with CTR, and subsequently captured on the microarray patterned with TIM4-Fc or streptavidin as pan-EV capture, as well as with 11 capture antibodies consisting of tetraspanins (CD82, CD81, CD63, and CD9), integrins (ITG- αV, ITG- α6, ITG-β1, and ITG- β5), and other markers (CD44, CD147 and ADAM10). The captured EV were then profiled for their expression of ITG-β1 and the $\mathrm{CoEn}_{\mathrm{Capture}^+}^{\mathrm{ITG-\beta1}}$ was calculated using $\mathrm{Ex}^{\mathrm{ITG\text{-}\beta1}}$ obtained from either TIM4 capture or streptavidin capture. As shown in Figure 3G, TIM4 and streptavidin capture were consistent for all capture-detection pairs tested here, affirming TIM4's usefulness as a pan-EV capture probe and in the determination of CoEn.

## 2.5 Validation of CoEn by Interfering with Syntenin-1-CD63 Binding in EVs

We used antibody microarrays with multiplexing of capture and detection antibodies to measure Ex and CoEx, and derive pairwise CoEn for multiple proteins using only 50 µL of sample. As discussed, CoEn of proteins may arise as a result of direct binding (i.e., protein complex formation), or simply because of sharing a common biogenesis pathway in the case of non-interacting proteins. Therefore, by re-routing a protein A away from its binding partner protein B during EV biogenesis, one would expect to reduce $\mathrm{CoEn}_{\mathrm{B}^+}^{\mathrm{A}}$. To evaluate whether experimentally measured $\mathrm{CoEn}_{\mathrm{B}^+}^{\mathrm{A}}$ could reveal re-routing, we selected two well-characterized proteins known to bind to one another during EV biogenesis, CD63 and Syntenin-1.[20–23] The tetraspanin CD63 is commonly found in EVs of exosomal origin and its C-terminus is known to bind the PDZ domain of Syntenin-1, a cytosolic protein involved in multiple cellular pathways, including exosome biogenesis. Syntenin-1 also binds to activated leukocyte cell adhesion molecule (ALIX), another important protein in exosome biogenesis via its three LYPX(n)L motifs.[20] We impeded the association of CD63 and Syntenin-1 by blocking the PDZ domain of Syntenin-1 with a cytopermeable peptide analogous to the C-terminal tail of CD63, Figure 4A,[24,25] with the expectation that the $\mathrm{CoEn}_{\mathrm{CD63}^+}^{\mathrm{Syntenin-1}}$ will decrease in the treated condition, and thus differential CoEn of treated vs untreated cells $\mathrm{dCoEn}_{\mathrm{CD63}^+}^{\mathrm{Syntenin-1}} < 0$. Furthermore, given the known binding between ALIX and Syntenin-1, we also expect $\mathrm{dCoEn}_{\mathrm{CD63}^+}^{\mathrm{ALIX}} < 0$. Conversely, one would not expect significant changes in the Ex nor the CoEn of proteins that do not interact with Syntenin-1, ALIX or CD63.

We treated immortalized kidney cell line 293T with the CD63 peptide analogue and a negative control consisting of a scramble peptide sharing the same amino acids in a different sequence. The harvested EVs were captured on an antibody microarray comprising 13 capture antibodies against tetraspanins (CD82, CD81, CD63, and CD9), integrins (ITG- αV, ITG-β1, and ITG- β5), EV biogenesis markers (LAMP1, LAMP2, and CD147), two surface makers (EpCAM and ADAM10), along with the pan-EV capture TIM4. We probed the co-localization of 8 detection proteins with fluorescent antibodies against 2 cytosolic targets (Syntenin-1 and ALIX) and 6 membrane targets (CD44, CD63, CD147, ITG-β1, EpCAM, and LAMP1), and derived Ex, CoEn and dCoEn. Reagent driven cross-reactivity between the capture and detection antibodies was tested and not observed, confirming the validity of our measurements, SI Figure 3.

The Ex of Syntenin-1 and ALIX, but not of other proteins, increased following addition of the CD63 peptide analogue to 293T cells, Figure 4B. Our Ex measurements are in accordance with Suárez *et al*. who also observed a slight increase of Syntenin-1 level in EVs of CD63-peptide-treated SK-MEL-147 cells.[24] $\mathrm{CoEx}_{\mathrm{CD63}^{+}}^{\mathrm{ALIX}}$ , $\mathrm{CoEx}_{\mathrm{CD63}^{+}}^{\mathrm{Syntenin-1}}$, $\mathrm{CoEn}_{\mathrm{CD63}^{+}}^{\mathrm{ALIX}}$ , and $\mathrm{CoEn}_{\mathrm{CD63}^{+}}^{\mathrm{Syntenin-1}}$ of both treatment conditions are shown in Figure 4C and 4D, respectively, and we note that CoEn reflects the expected protein re-routing, but not CoEx. Indeed, $\mathrm{CoEx}_{\mathrm{CD63}^{+}}^{\mathrm{ALIX}}$ was unchanged between the two conditions, while $\mathrm{CoEx}_{\mathrm{CD63}^{+}}^{\mathrm{Syntenin-1}}$ was increased in the CD63-peptide treated condition, which we ascribe to a baseline increase of $\mathrm{Ex}^{\mathrm{Syntenin-1}}$. As expected, $\mathrm{dCoEn}_{\mathrm{CD63}^{+}}^{\mathrm{ALIX}}$ = -0.23 and $\mathrm{dCoEn}_{\mathrm{CD63}^{+}}^{\mathrm{Syntenin-1}}$ = -0.13 were both negative as the result of the CD63 analogue interference, Figure 4D. The numeric results indicate that for each copy of Syntenin-1 depleted from $\mathrm{CD63}^{+}$ EVs, around 1.6 copies of ALIX were depleted along with it. The observation is explainable by the fact that Syntenin-1 has multiple ALIX binding sites,[20] and showcases the precision and sensitivity of CoEn measurements.

Beyond CD63, ALIX and Syntenin-1, we measured 104 pair-wise dCoEn of EVs from treated and control cells. Our custom heatmap in Figure 4E (see SI Supplementary Figure 4 for CoEn) visualizes these CoEn scores by color coding, with box heights representing statistical significance, scaled inversely to the p-value obtained under the null-hypothesis that $\mathrm{CoEn} = 0$. Given the large number of calculated CoEn scores, multiple testing corrections were applied to the p-values to account for false positives: boxes above minimum height indicate CoEn scores passing the common Benjamini-Hochberg threshold, while full-size boxes represent scores also passing the much stricter Bonferroni threshold. Interestingly, except for the expected changes in dCoEn of Syntenin-1 and ALIX with CD63, no other changes met statistical significance thresholds for any of $\mathrm{dCoEn}_{\mathrm{Capture}^+}^{\mathrm{CD44}}$, $\mathrm{dCoEn}_{\mathrm{Capture}^+}^{\mathrm{CD63}}$, $\mathrm{dCoEn}_{\mathrm{Capture}^+}^{\mathrm{CD147}}$, $\mathrm{dCoEn}_{\mathrm{Capture}^+}^{\mathrm{EpCAM}}$, $\mathrm{dCoEn}_{\mathrm{Capture}^+}^{\mathrm{ITG}-\beta1}$ and $\mathrm{dCoEn}_{\mathrm{Capture}^+}^{\mathrm{LAMP1}}$. These results are consistent with the high specificity of CD63 peptide interference, and the self-consistency between expected and measured findings suggests that dCoEn can serve as an accurate measure of protein re-routing.

Visually, the distribution of the CoEn values for EVs from CD63-peptide-treated and scramble treated cells look similar, but dCoEn (n = 29, minimum dCoEn = 0.17) reveals that there is a systematic down-regulation with the CD63 peptide, Figure 4F, with the majority linked to Syntenin-1 and ALIX. We observed a total of 7 EV subpopulations defined by their capture marker with significantly negative $\mathrm{dCoEn}_{\mathrm{Capture}^+}^{\mathrm{Syntenin}-1}$ and $\mathrm{dCoEn}_{\mathrm{Capture}^+}^{\mathrm{ALIX}}$, summarized in Figure 4G. Tetraspanins CD81 and CD9, lysosomal proteins LAMP1 and LAMP2 all had a negative dCoEn with both Syntenin-1 and ALIX. Interestingly, the degree of ALIX's depletion was more than 2-fold greater than of Syntenin-1's depletion in $\mathrm{CD81}^+$ and $\mathrm{CD9}^+$ EVs, mirroring the results observed for $\mathrm{CD63}^+$ EVs. Lastly, Syntenin-1 and ALIX dCoEn were also negative for $\mathrm{ITG}\text{-}\beta5^+$ and $\mathrm{EpCAM}^+$ EVs. Syntenin-1 can also bind to Syndecan, an EV cargo sorting protein, at its PDZ domains. Thus, the observed negative dCoEn while not anticipated, are explainable because integrins associate with Syndecan via the Syndecan-Syntenin-1-ALIX pathway,[22] and because impairment of Syntenin-1's PDZ domains was shown to impair sorting of EpCAM cargo into EVs.[26]

We validated key Ex, CoEx and CoEn patterns obtained by (ensemble) microarray analysis using SPFI with true single EV measurements. CD63-peptide-treated and scramble-treated 293T EVs were captured electrostatically on a poly-L-lysine-coated coverslip, and registered label-free with SPFI, thus circumventing the need for pan-EV staining, and freeing fluorescence imaging for antibody staining. The EVs were then stained for their marker expression of Syntenin-1, ALIX, CD63, and ITG-β1. Here, $\mathrm{Ex}^{\mathrm{A}}$ was derived by dividing the total A detection signal of $\mathrm{A}^+$ EVs by the number of total EVs analyzed, while $\mathrm{CoEx}_{\mathrm{B}^+}^{\mathrm{A}}$ was derived by dividing the total A detection signal of $\mathrm{B}^+$ EVs by the number of $\mathrm{B}^+$ EVs analyzed (see Section 2.1, Equations 2 and 3). SPFI-derived Ex measurements confirmed an increase in ALIX and Syntenin-1 expressions as observed in our ensemble microarray data (Figure 4B and SI Figure 5). Furthermore, SPFI-derived CoEn of Syntenin-1 and ALIX across EV subpopulations defined by CD63 and ITG-β1 revealed CoEn patterns consistent with our microarray measurement: negative dCoEn in $CD63^+$ EVs, while no changes in ITG $\beta 1^+$ EVs, as shown in SI Figure 6B and 7B. Recapitulating our microarray observation, CoEx failed to capture the expected changes at single EV level, SI Figure 6A and 7A.

In summary, CoEn for proteins targeted by the CD63-binding inhibition changed according to expectations, while several unanticipated changes in CoEn were uncovered that could be accounted for known protein associations and re-routing. Many protein pairs with significant depletion (i.e., dCoEn < 0) did not have a significant change in CoEx, nor Ex of the individual protein. We thus conclude that we have validated CoEn to be a sensitive and specific measure of protein co-enrichment in EVs.

## 2.6 CoEn of Proteins in EVs of 293T Cells Treated with Bafilomycin A1

Exosomes, which are an EV subtype of endosomal origin, are formed and matured within multivesicular bodies (MVBs) and secreted upon MVB fusion with the cellular plasma membrane. MVBs however can also fuse with lysosomes leading to the degradation of the EV cargo. Here, we exposed cells to Bafilomycin A1 (BafA1), a lysosomal acidification inhibitor known to prevent MVB-lysosome fusion,[12,27–29] to re-route vesicles destined for lysosomal degradation for secretion, illustrated in Figure 5A. Thus, one may expect that BafA1 treatment would alter the CoEn profiles for many protein pairs, measurable by dCoEn analysis between BafA1-treated and control EVs. In particular, one may expect the lysosomal-associated proteins 1 (LAMP1), which normally would be directed towards lysosomes, to be expressed more abundantly in exosomes. LAMP 1 has been shown to preferentially associate with $CD63^+$ EVs,[12] and thus we expect positive co-enrichment with CD63 and with other proteins and EV subpopulations as a result of BafA1 treatment.

293T cells were treated with either BafA1 or DMSO (solvent control), and the harvested EVs were profiled using microarrays with 13 capture antibodies and probed with 8 detection antibodies consisting of 3 cytosolic targets (Syntenin-1, ALIX, and HSP70) and 5 membrane targets (CD44, CD63, CD147, ITG-β1, and LAMP1). We evaluated whether BafA1 or DMSO might affect TIM4 EV capture by disturbing the lipidomic pathways and thus compromise its universality. We compared Ex of several proteins for EVs immobilized via TIM4 capture or streptavidin capture of biotinylated EVs and observed significant differences (SI Figure 8). Thus, we used streptavidin-capture to measure Ex and to calculate CoEn for this experiment.

We observed the expected increase in EV secretion following of BafA1 treatment, Figure 5B. Antibody-based capture EV subpopulations expressing CD44, LAMP1 and LAMP2 following BafA1 treatment all showed a higher CTR signal indicating increased subpopulation size (SI Figure 9). $\mathrm{Ex}^{\mathrm{LAMP1}}$ and $\mathrm{Ex}^{\mathrm{CD44}}$ also significantly increased upon BafA1 treatment, while the Ex of Syntenin-1 significantly decreased (Figure 5C).

In untreated 293T EVs, LAMP1 CoEn was strongly positive with two other lysosomal membrane glycoproteins, CD63 (i.e., LAMP3) and LAMP2, while it was negative with known ectosomal and membrane proteins such as CD9, CD81, CD147, and ITG-β1 (SI Figure 10) Interestingly, in DMSO-treated 293T EVs the dCoEn of CD9, CD81, CD147, and ITG-β1 in $\mathrm{LAMP1}^{+}$ and $\mathrm{LAMP2}^{+}$ EVs was positive compared to non-DMSO conditions, suggesting that DMSO alone may modulate EV cargo sorting of lysosomal proteins. BafA1 further increased CoEn of many proteins, and CoEn of LAMP1 was increased in $\mathrm{CD63}^{+}$, $\mathrm{CD9}^{+}$, $\mathrm{CD81}^{+}$, $\mathrm{CD147}^{+}$ and ITG- $\beta1^{+}$ EVs (Figure 5D). Interestingly, we observed a strong asymmetry between $\mathrm{dCoEn}_{\mathrm{Capture}^{+}}^{\mathrm{LAMP1}}$ and $\mathrm{dCoEn}_{\mathrm{LAMP1}^{+}}^{\mathrm{Detection}}$. dCoEn of all detection targets was negative in the $\mathrm{LAMP1}^{+}$ EV subpopulation, except for CD63. The observed asymmetry, together with the increase in the subpopulation of $\mathrm{LAMP1}^{+}$ and $\mathrm{LAMP2}^{+}$ EVs (SI Figure 9), are consistent with an increase of LAMP1 and LAMP2 secretion via EVs and larger numbers of $\mathrm{LAMP1}^{+}$ and $\mathrm{LAMP2}^{+}$ EVs.

While Ex of CD44 increased in response to BafA1 treatment, we observed both positive and negative dCoEn depending on the subpopulation. dCoEn of CD44 was positive for $\mathrm{CD147}^{+}$ and ITG-$\beta1^{+}$ EVs, but negative for $\mathrm{LAMP1}^{+}$, $\mathrm{ADAM10}^{+}$, ITG-$\beta5^{+}$ and $\mathrm{LAMP2}^{+}$ EVs. Our interpretation is that the additional CD44 secreted as a result of BafA1 treatment is selectively co-enriched depending on the EV subpopulation. Similar subpopulation selectivity was also observed in the case of Syntenin-1, where the Ex of Syntenin-1 decreased in BafA1-treated EVs, and $\mathrm{dCoEn}_{\mathrm{ITG-\beta1}^{+}}^{\mathrm{Syntenin-1}}$, $\mathrm{dCoEn}_{\mathrm{CD44}^{+}}^{\mathrm{Syntenin-1}}$, $\mathrm{dCoEn}_{\mathrm{CD147}^{+}}^{\mathrm{Syntenin-1}}$ were positive while $\mathrm{dCoEn}_{\mathrm{LAMP1}^{+}}^{\mathrm{Syntenin-1}}$ and $\mathrm{dCoEn}_{\mathrm{LAMP2}^{+}}^{\mathrm{Syntenin-1}}$ were negative. While BafA1 treatment did not significantly affect $\mathrm{Ex}^{\mathrm{CD63}}$, $\mathrm{Ex}^{\mathrm{ALIX}}$, $\mathrm{Ex}^{\mathrm{ITG-\beta1}}$, $\mathrm{Ex}^{\mathrm{CD147}}$ and $\mathrm{Ex}^{\mathrm{HSP70}}$, their dCoEn profile were greatly altered, highlighting both dCoEn's sensitivity in changes in cargo sorting, and by contrast, Ex's limitation in this regard. We thus find that CoEn shed light on numerous protein re-routing within EVs following treatment that are not visible by Ex or CoEx analysis.

## 2.7 *CoEn of Proteins in 293T, HT29 and MDA-MB-231 EVs*

EV biogenesis pathways have been known to shift with perturbations to homeostasis and pathogenic processes. Notably in cancer cells the EV biogenesis machinery is repurposed towards promoting tumor progression and metastasis.[11] We asked whether CoEn measurements could provide insight into the shared and distinct mechanisms of cargo sorting between EVs of cancer and non-cancer origin. We mapped the CoEn of up to 240 protein pairs in EVs derived from three distinct yet commonly used cell lines, including 293T immortalized kidney cells, HT29 colorectal cancer cells, and MDA-MB-231 triple-negative breast cancer cells. The EVs were profiled using microarrays with up to 15 capture antibodies (EpCAM, CD9, CD63, CD81, CD147, ADAM10, CD44, CD82, ITG- α6, ITG- β1, ITG-B4, ITG- αV, ITG- β5, LAMP1 and LAMP2) and probed with up to 16 detection antibodies consisted of 5 cytosolic targets (Syntenin-1, ALIX, TSG101, LC3 and HSP70) and up to 11 membrane targets (CD44, CD63, CD81, CD9, CD147, EpCAM, ITG- β1, ITG- β4, ADAM10, Syndecan and LAMP1).

The Ex and CoEn are shown in SI Figure 11 and Figure 6A and 6B and 6C. EVs from different cell lines showed significant variation in Ex (SI Figure 11), as well as variations in certain pairs of CoEn (Figure 6A and 6B and 6C). Most CoEn scores are neutral to positive in EVs of non-cancerous 293T. The more negative CoEn are found in EVs of cancer cell lines HT29 and MDA-MB-231 and could be interpreted as active protein separation. For example, in HT29 and MDA-MB-231 EVs, the integrin heterodimer ITG-α6 and ITG-β4 were consistently found to be negatively co-enriched with CD63, LAMP1, LAMP2, HSP70 and Syntenin-1. Notably, a number of CoEn were consistent across EVs derived from the three cells. We observed that LAMP1 is consistently positively co-enriched with itself, LAMP2 and CD63, and negatively co-enriched with CD9, CD81, CD82, CD147 and ADAM-10. We used unsupervised hierarchical clustering and principal component analysis (PCA) of pair-wise CoEn to identify pan-cell-line patterns and CoEn-based protein families. Interestingly, this analysis unveiled broad, pan-cell-line patterns, Figure 6D and 6E and 6F, that we interpret as EV protein co-packaging relationships. The clustering patterns based on the detection-defined CoEn patterns were surprisingly similar, irrespective of widely differing Ex and individual CoEn scores between the cell line EVs. We observed five consistent clusters of proteins based on PCA analysis: Group 1 consisting of LAMP1 (in yellow); Group 2 consisting of CD63, TSG-101, ALIX, and Syntenin-1 in 293T and HT29 EVs, or CD63 alone in MDA-MB-231 EVs (in red); Group 3 consisting of CD9, CD81, Syndecan, LC3 and EpCAM (in blue); Group 4 consisting of CD147 in 293T, or CD147 and ITG-β4 in HT29 and MDA-MB-231 EVs (in purple); and Group 5 consisting of CD44, ADAM10 and ITG-β1 (in green). We note that Group 1 consists of a prominent lysosomal protein, Group 2 of prominent exosomal proteins, while Group 3, 4 and 5 consist of membrane-residing proteins found in ectosomes. Interestingly, we observed different groupings for MDA-MB-231 cells where Syntenin-1 and ALIX clustered not with Group 2, but with Group 3 proteins. Likewise, CD147, an assumptive ectosomal protein, shifted from Group 4 to Group 5. We conclude that individual CoEn, as well as patterns of CoEn can help visualize and map co-packaging of EV proteins.

## 2.8 CoEn in EVs of Breast Cancer Cells with Known Organotropism

We further explore the potential of CoEn analysis for EV-based biomarker discovery with MDA-MB-231 breast cancer metastatic organotropism. MDA-MB-231 triple negative breast cancer cells are known for their aggressivity and propensity to metastasize to distant organs. Integrin expression in EVs has been linked to organotropism, and notably ITG-α6 and ITG-β4 have been associated with lung organotropism.[30] Here we profiled CoEn in EVs of parental MDA-MD-231 cells and in four sublines that were artificially selected for their metastasis to brain (831), bone (1833), liver (6133) and lung (4175), and that are each organotropic for their respective organ. We measured Ex (SI Figure 12), and CoEn (Figure 7A) of tetraspanins and cancer-associated proteins CD44 and EGFR, as well as the dCoEn (Figure 7B) of each organotropism relative to the parental line.

The Ex for the same targets remained relatively similar (SI Figure 12) regardless of variation in EV population size. Lung-tropic 4175 EVs had the most abundant ITG-β4 subpopulation (SI Figure 13), consistent with previous studies.[30] Meanwhile, the CoEn of tetraspanins CD9, CD63 and CD81 greatly differed in the ITG-$\beta4^+$ EV subpopulation of each organotropism (Figure 7A). ITG-$\beta4^+$ EVs negatively co-enriched with CD63 in all tropisms and positively co-enriched with CD9 in all except brain-tropism. CoEn of ITG-β4 and other detection targets was much more variable. CD81 and CD44 were positively co-enriched with ITG-$\beta4^+$ EVs in liver-tropism only, and negatively co-enriched with ITG-$\beta4^+$ EVs in lung- and brain-tropism; EGFR was positively co-enriched with ITG-$\beta4^+$ EVs in parental MDA-MB-231, and negatively in brain-tropism, and without significant co-enrichment in bone-, lung- and liver-tropism. For EVs of all tropisms except liver-tropic ones, ITG-β4's binding partner ITG-α6 followed the same co-enrichment trends.

CoEn of other EV subpopulations also varied significantly across tropisms, particularly among tetraspanins, CD82, CD44, EGFR and integrins. In brain tropism, integrin$^{+}$ EVs subpopulations consistently yielded negative co-enrichments for all detection targets, with the most negative being CD63. Overall, the CoEn patterns of brain-tropism were mostly negative, and more negative than other tropisms.

Next, we investigated the dCoEn of each organotropic subline relative to the parental MDA-MB-231 cell line gain insight into cargo-sorting changes specific for each organotropism, Figure 7B. Negative dCoEn values were more common than positive ones, with brain-tropism showing systemic downregulation of CoEn. Interestingly, the dCoEn patterns were distinct among the organotropisms with unique positive and negative dCoEn values. For example, $\mathrm{dCoEn}_{\mathrm{ITG-\alpha6^{+}}}^{\mathrm{CD81}}$ and $\mathrm{dCoEn}_{\mathrm{ITG-\beta4^{+}}}^{\mathrm{CD81}}$ were positive in liver-tropism only, meanwhile, negative $\mathrm{dCoEn}_{\mathrm{ITG-\beta4^{+}}}^{\mathrm{CD44}}$ was only seen in lung-tropism. The greatest tropism-specific positive dCoEn were $\mathrm{dCoEn}_{\mathrm{ITG-\alpha V^{+}}}^{\mathrm{CD44}}$ and $\mathrm{dCoEn}_{\mathrm{ITG-\alpha V^{+}}}^{\mathrm{CD63}}$ in the EVs of the bone-tropic cells with an 8-fold and 4-fold increase compared to parental EVs, respectively. While preliminary, our results suggest that CoEn and dCoEn reveal differences between EVs from MDA-MB-231 and its organotropic sublines, highlighting their potential as a basis for developing CoEn-based biomarkers.

## 3. Discussion:

### *3.1 CoEn is Consistent with, and Offers Insights into, EV Biology*

In this work, we introduced a CoEn as quantitative co-enrichment metric which lends itself to high throughput subpopulation analysis along with dCoEn to measure relative co-enrichment changes. We quantified Ex, CoEx and CoEn here using a custom antibody microarray, with key Ex, CoEx, CoEn signatures further validated with single EV measurements using SPFI.[17] Our results suggest that both ensemble assays such as microarray, ELISA and bead-based assays[31–34] and single EV methods are suitable for measuring Ex, CoEx and CoEn. To support the measurement and adoption of CoEn, we provide a technical workflow to derive Ex, CoEx and CoEn using conventional ELISA, as detailed in the Supplementary Method.

Unlike Ex and CoEx, CoEn is a dimensionless, normalized metric which reliably detects non-random protein association on EVs. It is independent of the individual protein abundance and thus insensitive to baseline changes in Ex between experimental conditions. Due to their intrinsic normalization, CoEn values are furthermore well-suited for hierarchical clustering and PCA analysis, effectively achieving the same function as typical data preprocessing steps such as log transformation and z-score standardization.

We hypothesized that proteins with fully shared biogenesis pathways yield highly positive CoEn with each other and highly correlated CoEn patterns, while proteins associated with different EV biogenesis pathways would have a negative CoEn. We validated CoEn using a peptide-induced re-routing of Syntenin-1 away from its binding partner CD63 as well as a BafA1 treatment. Each revealed the expected loss and gain of association in co-enrichment, while also revealing new associations, many of which could be explained based on known EV biology. Clustering and PCA uncovered that CoEn of various protein families in EVs from 293T, HT-29 and MDA-MB-231 were conserved, despite differences in Ex. Proteins with known binding or cargo sorting relationships (e.g., ALIX, Syntenin-1 and CD63) clustered together based on their CoEn patterns (Figure 6). Meanwhile, lysosomal protein LAMP1 was consistently positively co-enriched with other purported exosomal proteins such as CD63, ALIX and Syntenin-1, while being negatively co-enriched with membrane proteins such as CD81, CD9, CD147 and EpCAM. As a result, LAMP1, which was co-depleted with other membrane-residing proteins, was clustered with prominent purported ectosomal markers on the opposite end of the PCA graph. (Figure 6E and 6F and 6G). Again, expected changes were observed in CoEn, but not reflected in Ex or CoEx measurements (SI Figure 11 and 14). Interestingly, we observed that in 0.1 % DMSO-treated 293T EVs, LAMP1's negative co-enrichment with membrane proteins was attenuated (SI Figure 10). Given DMSO's known membrane-disruptive properties,[35] our result hints at the potential disturbance of DMSO in EV biogenesis and thus caution should be applied when using DMSO as a solvent control in EV studies.

In addition to lysosomal Group 1 (LAMP1) and exosomal Group 2 (CD63, ALIX and Syntenin-1) protein families, CoEn identified three ectosomal (e.g., plasma membrane) protein families: Group 3 CD9, CD81, Syndecan, LC3 and EpCAM in Group 3; CD147 and ITG-β4 in Group 4; and CD44, ADAM10 and ITG-β1 in Group 5 (Figure 6D and 6E and 6F), revealing heterogeneity in the biogenesis of membrane-budded EVs. Some protein family members, such as Alix, Syntenin-1 and CD63 from Group 2 are well known for their co-packaging in exosomes due to direct interactions. Hierarchical clustering grouped Syndecan with EpCAM, a potential cargo of Syndecan in a Syntenin-dependent manner as previously shown.[26] Interestingly, we observed a shift of Syntenin-1 and ALIX from clustering with exosomal marker CD63 in Group 2 in 293T and HT-29 EVs, to clustering with Syndecan and EpCAM in Group 3 in MDA-MB-231 EVs, hinting at potential alteration in their cargo sorting mechanisms from CD63-Syntenin-ALIX dominant biogenesis route in 293T and HT-29 to a Syndecan-Syntenin-ALIX dominant biogenesis route in MDA-MB-231 cells. Other protein family members had not previously been shown to co-package in EVs, and CoEn may thus be the first direct evidence of such relationships. For example, while the direct interaction between Group 3 proteins CD9 and LC3 has not been reported, knock-out of CD9 has been shown to drastically reduce LC3-positive late endosomes *in vitro*,[24] hinting at a possible LC3-CD9 dependent EV biogenesis route. Similarly, while Group 4 proteins ITG-β4 and CD147 have not been linked within the EV context, CD147 knock-down led to significantly decreased ITG-β4 in cell line model.[36] Together, our results suggest that co-enrichment analysis could help distinguish and classify cargo-protein-defined EV biogenesis pathways including lysosomal, exosomal and ectosomal, and visualize variations and shifts in cargo-sorting.

## *3.2 Pan-EV Capture with Phosphatidylserine*

We benchmarked and validated phosphatidylserine-binding probes, namely Annexin V and TIM4 as a suitable pan-EV capture agent and found PS to be highly expressed in individual EVs on the outer leaflet of both cancer and non-cancer cells among the four cell lines tested here (i.e., IMR-90, 293T, HT29 and MDA-MB-231). Additional studies are needed to confirm our findings and to uncover mechanism of PS expression on the outer leaflet. We observed that the presence of DMSO, a molecule known to cause membrane deformities,[35] led to a discrepancy in measurements between 293T EVs captured by TIM4 and biotinylated EVs captured by streptavidin, which we ascribe to an alteration in PS expression on the EVs (SI Figure 8). Caution is advised when evaluating the use of TIM4 capture of $PS^+$ EVs in non-homeostatic conditions in which lipid pathways may be affected, and to confirm pan-EV expression of PS in the outer leaflet with independent methods. Other approaches to pan-EV capture may be useful such as lipid insertion probes or curvature-sensing peptides.[37,38]

## 3.3 CoEn May Not Reflect All Cases of Co-enrichment

We recognize that protein co-enrichment in EVs is multi-faceted, and metrics with alternative definitions of co-enrichment could offer complementary insights to the concentration-defined CoEn introduced in this work. For example, as detailed in Section 2.5, SPFI, a single EV method, was used to validate concentration-defined CoEn measured by antibody microarray, an ensemble assay. However, calculating concentrations from SPFI requires averaging single EV measurements to emulate ensemble measurements, inherently losing the frequency information through this operation. As a result, our current definition of $\mathrm{CoEn}^{\mathrm{A}}_{\mathrm{B+}}$ which is based on an ensemble measurement of protein A, cannot distinguish whether protein A is highly expressed in a few $B^+$ EVs, or present at low levels across many $B^+$ EVs. It is thus natural to conceive a CoEn definition based on the single EV expression frequency of A (i.e., by measuring the percentage positivity of $A^+$ EVs), to measure whether the frequency of $A^+$ EVs among $B^+$ EVs is increased or reduced compared to the overall frequency of $A^+$ EVs in the sample. We note that a frequency-based co-enrichment could extend pairwise CoEn analysis to a full combinatorial one, i.e., co-enrichment of three and more proteins could be detected simultaneously in single EVs. The SPFI validation experiments and emerging single EV methods[17,39] support the feasibility of frequency based CoEn analysis, provided that the measurement methods achieve sufficiently high detection sensitivities for each protein and support multiplex analysis.

## 3.4 Ex, CoEx, and CoEn Are Susceptible to Methodological Bias

The accuracy and precision of the experimentally-derived Ex, CoEx and CoEn values are subject to the processes and technologies used. In the case of antibody microarray, the microarray fabrication process,[31] the binding properties of the capture antibody, spatial bias across the microarray slide,[40] the brightness of the conjugated fluorophore, as well as the sensitivity and linearity of the microarray scanner could all influence the range and sensitivity of CoEn. While in theory CoEn can adopt the full range of negative and positive values, positive CoEn is favored. Negative CoEn means lower fluorescent signals, but in some cases these signals will fall below the detection threshold, and hence will not be reported, and thus prevent the computation of expected negative CoEn values. Furthermore, Ex and CoEx are subject to biases stemming from variation of detection antibody binding properties (i.e., a given RFU signal produced from anti-A antibody does not equal to the same signal produced from an anti-B antibody), making them not directly comparable between different detection targets. However, whenever the detection signal is above the detection threshold, CoEn is not affected by said detection antibody variation because this variation is expected to impact CoEx and Ex equally, and to be offset when CoEx and Ex are ratioed to calculate CoEn. Thus, CoEn values are expected to be comparatively invulnerable to experimental variation, and comparable among different detection targets, experiments and labs. Nonetheless, we note that the binding properties of the capture antibody may influence the measured CoEn values, as affinity of the capture antibody (and of a universal binder) could directly influence the amount of EVs captured, and low-affinity antibody or low concentration of a target can lead to scarce EV capture, and protein expression could then fall below the detection threshold of the universal detection dyes.

While not discussed explicitly, the number of proteins A and B co-expressed in individual EVs is expected to raise quickly with the size of the EVs simply because large EVs have the cargo capacity to carry more proteins. As $\mathrm{CoEn}_{\mathrm{B+}}^{\mathrm{A}}$, collapses the expression of B into a yes/no value (i.e., $B^+$), our approach is susceptible to produce high CoEn values for large EVs, were it not for normalization. Here we normalized using CTR dyes (i.e., volume) and streptavidin-biotinylation (i.e., surface area) that scale to the 3rd and 2nd power with diameter, respectively,[17] and which underpin the precision and accuracy of CoEn measurements. Self-CoEn, as in $\mathrm{CoEn}_{\mathrm{A+}}^{\mathrm{A}}$, was measured as a consequence of the combinatorial microarray design, but not given weight in the interpretation of our experiments. Indeed, one must consider the depletion of targets by capture antibodies that are expected to bias the number of detection antibodies that can bind, and thus yield a potential underestimate of CoEn.

We note that the EV isolation method can introduce size and/or proteomic biases.[41,42] In this work, size exclusion chromatography was used to isolate EVs from the collected cell supernatant, which inherently enriches EVs within a certain size range. Consequently, the reported Ex, CoEx, and CoEn measurement reflect this EV population and may be biased relative to the full EV population. The microarray analysis uses antibodies for capture and has been used with plasma, and is also compatible with unpurified cell supernatant (unpublished). Hence it would be possible to repeat this study with unpurified supernatant and compare the results obtained with purified EVs using different isolation methods to evaluate potential biases due to sample processing.

### *3.5 Potential of CoEn as Biomarker*

We explored CoEn's utility in identifying specific features in organotropic metastatic breast cancer cell line models that may serve as biomarkers. We found that CoEn patterns of EVs from a parental breast cancer cell line MDA-MB-231 and its several organotropic sublines were distinct, with tropism-specific CoEn patterns observed across all subtypes. For example, we found that ITG-β4, a beta integrin implicated in lung-tropic breast cancer metastasis, is selectively co-enriched (e.g., with CD9 and CD81) and co-depleted (e.g., with CD63 and CD44) in an EV-subpopulation-specific and tropism-specific manner. The sensitivity of CoEn and the specificity of CoEn pairs to specific organotropism within the same cell line suggest it could serve as a biomarker for disease diagnosis and classification. Future studies, notably CoEn analysis with more clinically relevant samples (i.e., patient samples) will be needed to fully assess CoEn's discriminatory power. We note that given the complex nature of biofluids, key assay components such as the pan-EV capture probe should be validated again. We also note that while microarray and other multiplex platforms are essential in the early stage of biomarker exploration, an eventual clinical CoEn assay would ideally only need to probe for the global expression and the CoEn pair of interest, with ease of use and sample throughput as main priorities.

## 4. Conclusion

Here, we introduced and mathematically defined the CoEn and dCoEn of proteins in EVs. Furthermore, we introduce an experimental framework to assess Ex, CoEx, and CoEn of proteins in EVs using antibody microarrays. We then measured and biologically validated CoEn and dCoEn by perturbating well known EV biogenesis pathways, one by interfering with syntenin-CD63 binding, and one using Bafilomycin A1 known to re-route intracellular vesicles away from lysosomes to exosomal secretion. CoEn and dCoEn were consistent with expectations, while also uncovering unexpected, yet often, logical changes. CoEn profiles of EVs were analyzed by PCA which produced distinct clusters of proteins that were conserved across cell lines and separating putative lysosomal, exosomal and ectosomal proteins. CoEn analysis of organotropic breast cancer cell lines revealed differences in CoEn patterns, suggesting a potential use as biomarker. Collectively, our results showed that CoEn uncovers information not measurable by Ex and CoEx, and that it could reflect and reveal physiological processes, notably the ones linked to EV biogenesis, in health and disease. Hence CoEn may be used to infer EV biogenesis, and generate hypothesis, simply based on protein associations within EVs, thus providing simpler alternative to direct, and technically challenging, observations of biogenesis events, as well as uncover new, or confirm existing, pathways.

CoEn and dCoEn analysis could be extended to include nucleic acids, lipids and other molecular cargo contained within EVs, and applied across different types of molecular cargo, e.g., the co-enrichment of an RNA cargo A in $B^+$ EV subpopulations. We expect the CoEn workflow to be transferrable to various bulk assays, including common bead-based assays,[31–34] and to single EV analysis.[17,39] The study and definition of EV phenotypes and of molecular cargo co-packaging in EVs; broader efforts to establish health and disease states; the discovery of biomarkers for disease diagnosis and monitoring; as well as characterization and quality control of EV-based therapies, all stand to gain from CoEn analysis.

## Materials and Methods:

### Cell Culture

Immortalized embryonic kidney cell line 293T, primary lung fibroblast cell line IMR-90, human colorectal cancer cell line HT29, triple negative breast cancer cell line MDA-MB-231 and its organotropic sublines 1833, 4175, 6113, and 831 were cultured in Dulbecco's Modified Eagle Medium (Gibco, USA) containing 4.5 g/L D-glucose, 4.5 g/L L-glutamine, and 110 mg/mL sodium pyruvate. The media was further supplemented with 10% FBS and 1% penicillin-streptomycin (Gibco, USA). All cell lines were incubated at 37° C with 5% $CO_2$ supplementation.

### EV Isolation

Cells were grown in tissue culture flasks until 50% confluency, after which the culture media was replaced with DMEM media containing 5% EV-depleted FBS (or any specified media), and collected after 48 h. The media was centrifuged at 400×g for 15 min, syringe-filtered using a 0.22 µm PES membrane filter (Millipore Sigma, USA), and then concentrated using an Amicon Ultra 100 kDa NMWCO centrifugal filter unit (Millipore Sigma, USA). The media was concentrated down to 500 µL per every 45 mL raw supernatant and purified using 70 nm qEVoriginal Columns (IZON, USA) with PBS used as the loading buffer. The column was flushed with 20 mL of PBS, loaded with 500 µL of concentrated EV-containing supernatant and ten 500 µL fractions were collected. Fraction 7 to fraction 10 were pooled for further experiments. Depending on its concentration, the purified EV solution may be further concentrated using Pierce™ PES 100 MWCO 0.5 mL Protein Concentrators (Thermofisher Scientific, USA).

### SPFI Workflow

Purified cell line EVs were immobilized on poly-L-lysine (PLL) functionalized coverslips and label-free EV sizing and fluorescence measurements were acquired as described previously.[17] For PLL functionalization, glass coverslips were first sonicated isopropyl alcohol for 5 min sonicated in distilled water for 1 min, blow-dried with nitrogen, and plasma treated for 3 min. Plasma-activated coverslips were then incubated with 0.1% PLL in Milli-Q water for 30 min and rinsed with PBS. EVs were incubated on the PLL-functionalized coverslip and binding monitored in real-time using SPFI until a desired capture coverage has been reached, usually 1 to 5 min. Captured EVs were first imaged for particle registration and sizing before additional staining. For on-slide

PS staining, FITC-conjugated Annexin V (Biolegend, USA) was diluted 1: 50 in Annexin V binding buffer (Biolegend, USA) and incubated for 1 h. For antibody immunostaining, oligo-conjugated antibodies were incubated at 5 µg/mL for 2 h diluted in PBS with 3% BSA, sheared salmon sperm DNA (Thermo Fisher, USA) and 0.5% dextran sulfate (Sigma, USA), and additionally with 0.05% Saponin for cytosolic targets. Stained EVs were imaged again and marker positive EVs were quantified using our previously published, openly available imaging processing pipeline (https://github.com/junckerlab/SPFI).

**Peptide and Bafilomycin A1 Treatment**

Membrane-permeable peptides with the sequence RRRRRRRCCLVKSIRSGYEVM (CD63) and RRRRRRRSMGYESLSIVRSVK (scrambled control) were synthesized by LifeTein, USA. Immortalized embryonic kidney cell line 293T were treated with 10 µM of CD63- or scrambled peptide for 48 h. We validated that this 48 h treatment duration was sufficient, as an extended 120 h treatment yielded comparable results (SI Figure 6C). For Bafilomycin A1 treatment, immortalized embryonic kidney cell line 293T was treated with 100 nM Bafilomycin A1 (Abcam Inc, USA) or DMSO (solvent control) for 48 h. The supernatants were collected for downstream EV isolation via SEC.

**Fluorescent EV Labeling with CFDA-SE and Cell Tracker Red**

Carboxyfluorescein diacetate succinimidyl ester (CFDA-SE, Thermofisher Scientific, USA) was added to purified EV solution at a concentration of 40 µM and incubated for 2 h at 37º C. Cell Tracker Red (Thermofisher Scientific, USA) was added to purified EV solution at a concentration of 40 µM and incubated for 2 h at room temperature. Excess dye was removed using Pierce™ PES 100 MWCO 0.5 mL protein concentrators.

**EV Biotinylation**

EZ-Link NHS-LC-LC-Biotin (Thermo Scientific, USA) was added to the purified EV solution to achieve a final biotin concentration of 0.1 mg/mL, 0.5 mg/mL or 1 mg/mL and incubated at room temperature for 30 min Pierce™ PES 100 MWCO 0.5 mL protein concentrators were used to remove unreacted biotin reagent.

**Antibody Microarray Fabrication**

All antibodies used in this work can be found in Supplementary Table 1. Antibodies were diluted in spotting buffer containing 15% 2,3-butanediol and 1 M betaine in PBS to achieve a final concentration of 100 µg/mL for the organotropic MDA experiment and 300 µg/mL for all other experiments. TIM4 and streptavidin were diluted in the abovementioned buffer to a final concentration of 300 µg/mL. The solutions were loaded into an Axygen 384-well polypropylene skirted polymerase chain reaction microplate (Fisher Scientific, USA) and the antibody microarray was fabricated onto 2D-Aldehyde glass slides (PolyAn, Germany) using a sciFLEXARRAYER SX (Scienion, Germany) equipped with a single piezo dispense capillary nozzle with coating 1 (Scienion, Germany). A relative humidity of 65% was maintained throughout the fabrication process and the printed microarrays were incubated overnight in the dark in a 70% humidity bell jar.

**Antibody Microarray Workflow**

The antibody microarray slides were washed in PBS with 0.1% Tween-20 (0.1% PBST) on a rotary shaker at 450 RMP for 15 min and then blocked with 0.1% PBST supplemented with 3% BSA for 2 h s at room temperature. After blocking, the slides were dried using a centrifugal slide dryer (Arrayit, USA) and assembled with a 16-well (or 64-well) microarray hybridization chamber (GRACE BIO-LAB, USA). 50 µL purified EV samples (25 µL for 64-well) were loaded to the designated wells and incubated at room temperature 2 h, then incubated overnight at 4C. After sample incubation, the wells were washed with PBS for 3 times on shaker and detection antibodies were loaded at a concentration between 1 µg ~ 5 µg/mL, diluted in 0.03% PBST with 3% BSA. The slides were incubated at room temperature on a shaker for 2 h. The detection antibodies were detected using either anti-species fluorescently-labelled secondary antibodies, fluorescently labelled streptavidin at a final concentration of 1 µg/mL diluted in 0.03% PBST with 3% BSA, or fluorescently labelled DNA oligos at a final concentration of 10 nM in PBS with 0.025% Tween-20, 300 mM NaCl, 0.5 mg/mL sheared salmon sperm DNA (Thermo Fisher, USA) and 0.5% dextran sulfate (Sigma, USA).[31] For intravesicular targets, an additional fixation (10 min with 4% paraformaldehyde), antigen retrieval (1 min with 90°C PBS), and quenching step (2 h with 0.03%PBST, 3% BSA and 0.3M glycine) were performed prior to detection antibody incubation.[31]

Finally, the slides were washed in a 0.03% PBST bath for 15 min (or 5 min for oligo-detected targets) on a shaker, dried using a centrifugal slide dryer before imaging. A microarray scanner (InnoScan 1100 AL, Innopsys, France) was used to visualize the microarray slides, where the gain and laser power were made consistent across the same experiment. ArrayPro 4.5 (Meyer Instruments, USA) was used to quantify the mean relative fluorescence unit (RFU) value for each microarray spots. Ex and CoEx were calculated using the RFU intensity values for each respective measurement and dye according to the workflow shown in Figure 2.

# Co-Enrichment of Proteins in Extracellular Vesicles

Molly L. Shen[1-4], Andreas Wallucks[1,2], Rosalie Martel[1,2,†], Zijie Jin[1,2,†], Lucile Alexandre[1,2], Philippe DeCorwin-Martin[1,2], Lorenna Oliveira Fernandes de Araujo[1,2], Andy Ng[1,2], Peter M. Siegel[3,4], and David Juncker[1,2,*]

[1]Biomedical Engineering Department, McGill University, Montréal, QC, Canada

[2]Victor Phillip Dahdaleh Institute of Genomic Medicine, McGill University, Montréal, QC, Canada

[3]Goodman Cancer Institute, McGill University, Montréal, QC, Canada

[4]Department of Medicine, McGill University, Montréal, QC, Canada

[†] R.M. and Z.J. contributed equally.

# Main Figures

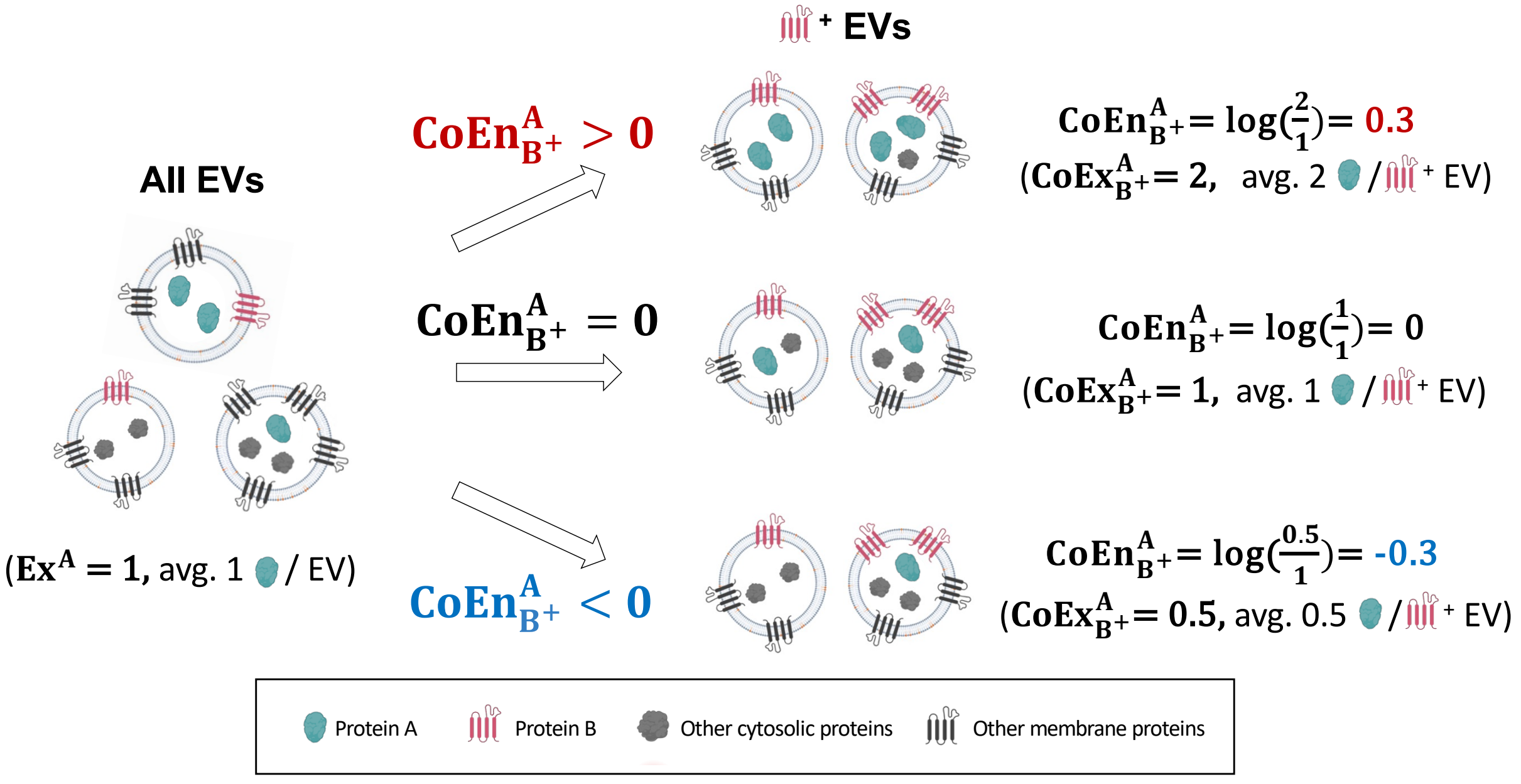

B

Biogenesis Pathway 1

Multivesicular body (MVB)

Biogenesis Pathway 2

Protein A
Protein B
Protein C
Protein D

Protein A and B are secreted together in biogenesis pathway 1:

$\mathrm{CoEn}^{A}_{B^+} > 0$ $\quad$ $\mathrm{CoEn}^{B}_{A^+} > 0$

Protein C is associated with a different biogenesis pathway than A and B:

$\mathrm{CoEn}^{C}_{A^+} < 0$ $\quad$ $\mathrm{CoEn}^{A}_{C^+} < 0$

$\mathrm{CoEn}^{C}_{B^+} < 0$ $\quad$ $\mathrm{CoEn}^{B}_{C^+} < 0$

Protein D is found in all EVs:

$\mathrm{CoEn}^{A,B,\mathrm{or}\ C}_{D^+} \approx 0$ $\quad$ $\mathrm{CoEn}^{D}_{A,B,\mathrm{or}\ C^+} \approx 0$

**Figure 1. Schematics of positive, neutral, and negative co-enrichment (CoEn) of proteins in EVs**. **(A)** CoEn of a protein A in EVs expressing protein B ($\mathrm{CoEn}^{A}_{B^+}$) is the logarithm of the ratio of (i) co-expression of a protein A in EVs expressing protein B ($\mathrm{CoEx}^{A}_{B^+}$) and of (ii) the global expression of protein A in all EVs ($\mathrm{Ex}^{A}$). $\mathrm{Ex}^{A}$ is first measured (left), and then $\mathrm{CoEn}^{A}_{B^+}$ calculated upon measuring $\mathrm{CoEx}^{A}_{B^+}$ (right). If loading of protein A into the EVs was a purely random process, $\mathrm{CoEx}^{A}_{B^+}$ and $\mathrm{Ex}^{A}$ would statistically be equal and hence CoEn = 0. Therefore, CoEn ≠ 0 can be seen as an indicator of non-random association (i.e., neither co-enrichment or co-depletion) of A and B: positive CoEn values ($\mathrm{CoEn}^{A}_{B^+}$>0, annotated in red) indicate that protein A is enriched in the B$^+$ EV subpopulation, while negative CoEn values ($\mathrm{CoEn}^{A}_{B^+}$ <0, annotated in blue) indicate that protein A is depleted in B$^+$ EVs. For simplicity, the number of EVs and size is kept equal in this schematic and thus Log(CoEx) and CoEn are proportional or even equal, but in actuality they are not. **(B)** Schematics showing hypothetical scenarios of pairwise CoEn combinations for 4 proteins in a cell with only 2 EV biogenesis pathways, namely protein A and B that are preferentially associated with pathway 1, protein C which is associated with pathway 2, and protein D which is present in both pathways. As a result of their associations, protein pair A and B are co-enriched while co-depleted with protein C. Meanwhile, protein D is neither co-enriched nor co-depleted with A, B, nor C. In reality, a pair of proteins can share multiple pathways together, making their measured CoEn a reflection of their overall co-enrichment relationships.

# Experimental Derivation of CoEn using Antibody Microarray

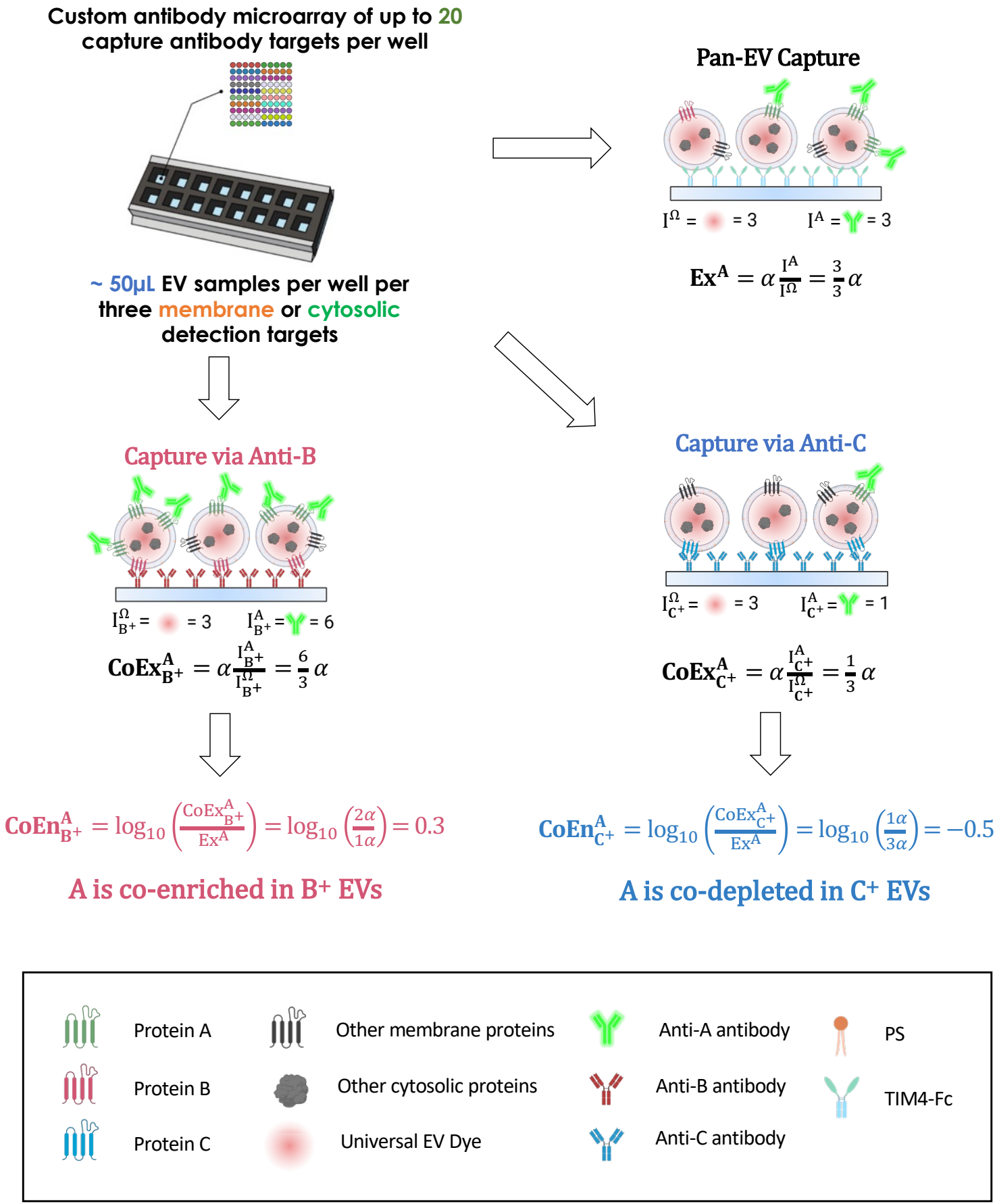

**Figure 2. Schematic of the experimental workflow and calculation of Ex, CoEx and CoEn using antibody microarrays**. EVs are captured either via a surface bound pan-EV capture (e.g. TIM4 that binds PS), or via antibody that binds EVs expressing protein B or protein C on their surface. The concentration of immobilized EVs is measured via the fluorescence intensity of the universal dye Ω (e.g. CTR dye), and the concentration of protein A in the EVs is quantified via a fluorescently labelled anti-protein A detection antibody. $Ex^{A}$ is calculated by normalizing the fluorescence of A (i.e. $I^{A}$) with the fluorescence of Ω for EVs captured with the pan-EV capture (i.e. $I^{\Omega}$). α is the normalization factor between $I^{A}$ and $I^{\Omega}$. Likewise, the co-expression of protein A in EVs captured with anti-B antibodies ($CoEx^{A}_{B+}$) and anti-C antibodies ($CoEx^{A}_{C+}$), respectively, is also calculated by normalizing the fluorescence of $I^{A}_{B+}$ with $I^{\Omega}_{B+}$ and $I^{\Omega}_{C+}$, thereby ensuring comparability. Finally, $CoEn^{A}_{B+}$ and $CoEn^{A}_{C+}$ are calculated by normalizing $CoEx^{A}_{B+}$ and $CoEx^{A}_{C+}$ with $Ex^{A}$. In the two examples shown here, $CoEn^{A}_{B+} > 0$ in $B^{+}$ EVs (on the left) indicates that A is co-enriched in $B^{+}$ EVs and conversely, a negative $CoEn^{A}_{C+} < 0$ in $C^{+}$ EVs (on the right) indicates that A is co-depleted in $C^{+}$ EVs. In this illustration the number and size of EVs in each scenario was kept equal for clarity.

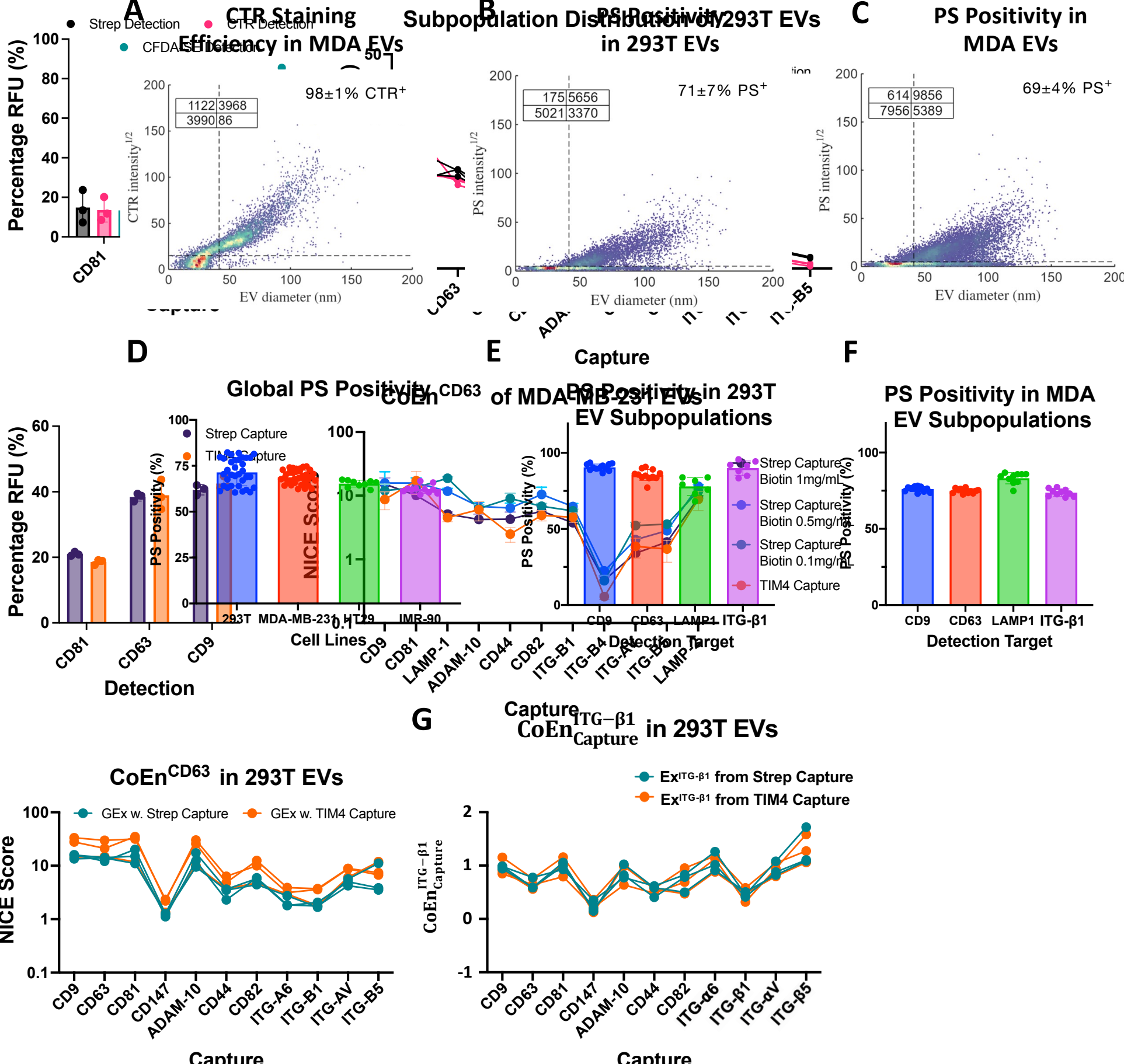

**Figure 3. Experimental validation of CTR as universal EV label and PS-binding probes Annexin and TIM4 as universal EV capture**. **(A)** EVs from MDA-MB-231 cells were dyed with CTR and SPFI used for label-free sizing of individual EVs and concomitant fluorescent detection of CTR, showing reliable detection and labeling of single EVs. **(B-C)** Representative SPFI scatterplot of EVs from **(B)** 293T and **(C)** MDA-MB-231 cells stained with Annexin-FITC that selectively binds to Phosphatidylserine (PS). Around 70% of EVs were PS+ in each of the cell lines. **(D)** Quantification of PS positivity in 293T, MDA-MB-231, HT-29 and IMR-90 cell line EVs showcasing that PS is highly abundant in cancer and non-cancer cell line EVs. For 293T and MDA-MB-231 EVs the number of biological replicates N=3 and for HT-29 and IMR-90 EVs N = 1 with the number of technical repeats n = 8 - 12 technical replicates, and ~10,000 single EVs analyzed per N. **(E)** 293T and **(F)** MDA-MB-231 EVs captured via physisorption on a poly-L-glycine-functionalized coverslip slide and probed for the expression of PS using Annexin-FITC and the expression of CD9, CD63, ITG-β1 and LAMP1 using detection antibodies, and >74% for each antibody-defined subpopulation were positive for PS (N = 1, n = 8 for LAMP1 and ITG-B1, n = 12 for CD9 and CD63). **(G)** CoEn of ITG-β1 in 11 293T EV subpopulations were calculated using antibody microarrays, while using TIM4 or streptavidin (0.1 mg/mL biotinylation concentration) as universal capture to calculate Ex and normalize signals. Data shows 3 biological replicates with 5 technical repeats each (N = 3, n = 5). In summary, the high CTR and PS positivity, the consistency across cell lines and different detection targets, and consistent results between TIM4 and streptavidin suggest broad PS expression and the use use of CTR and TIM4 as universal detection and capture, respectively.

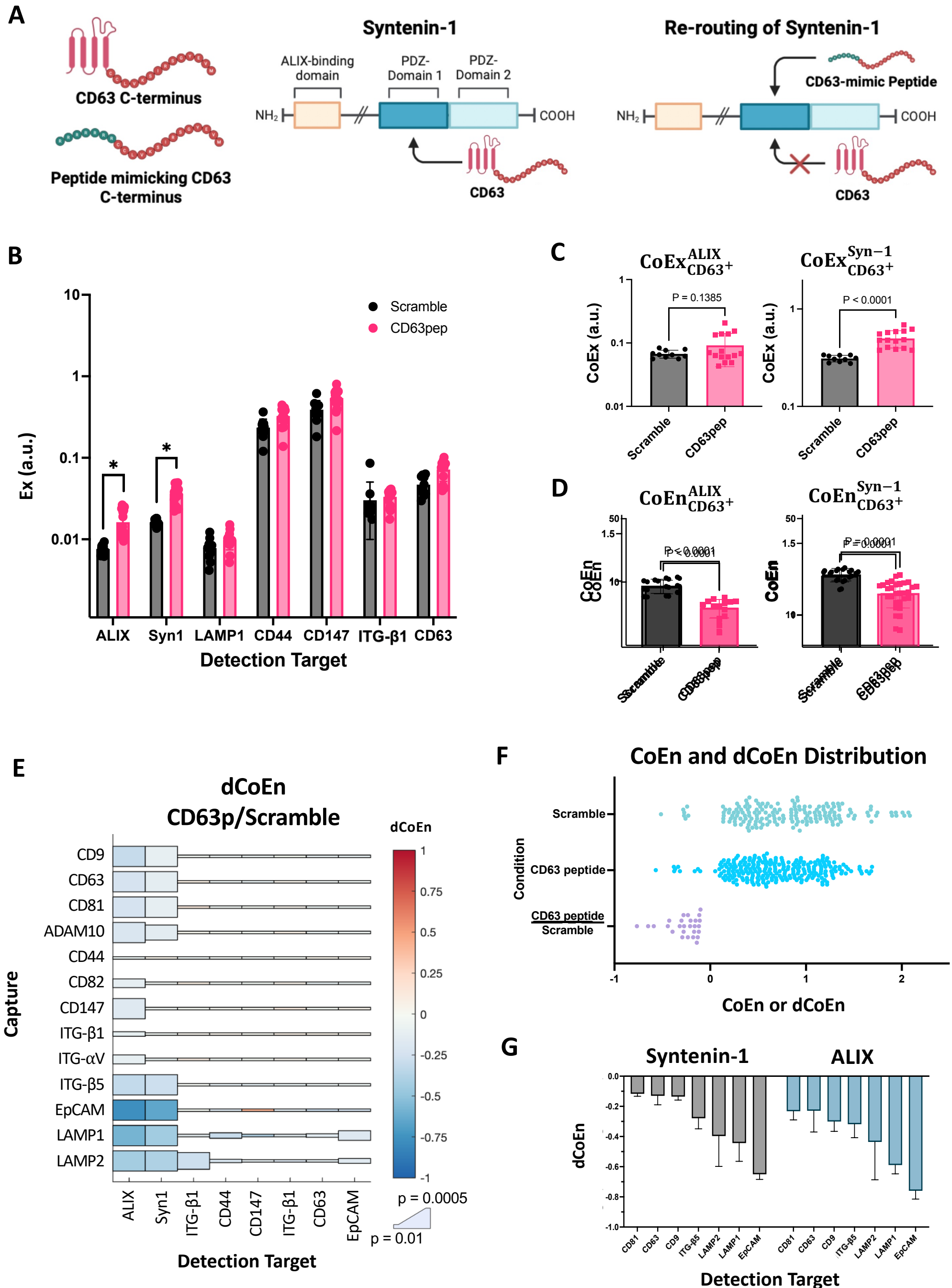

**Figure 4. Biological validation of CoEn and dCoEn via a peptide-induced blockade of Syntenin-CD63 binding**. 293T cells were treated with 10 μM of CD63-blocking or scramble peptide for 48h. **(A)** Summary schematic of peptide-induced blockade of Syntenin-CD63 binding. **(B)** Ex values of all detection targets. (biological repeat N = 3, technical repeat n = 5 for blocking peptide treated condition, N = 2, n = 5 for scramble peptide treated condition). p-Value cutoff of p = 0.0005 was determined using the Bonferroni correction. **(C)** CoEx and **(D)** CoEn of ALIX and Syntenin-1 in CD63$^{+}$ EVs in CD63- or scrambled peptide treated 293T. **(E)** Heatmap of all 106 dCoEns of EVs secreted following CD63-blocking-peptide relative to the scramble-peptide condition. Lower significance cutoff of -Log(p-Value) = 1.97 (p = 0.01) and upper significance cut-off -Log(p-Value) = 3.29 (p = 0.0005) defined based on Benjamini-Hochberg and Bonferroni correction procedures, respectively (N = 3, n = 5 for CD63-blocking peptide and N = 2, n = 5 for scramble peptide treated conditions). **(F)** Scatter plot of CoEn for CD63-blocking-peptide and scramble-peptide treated 293T EVs, and dCoEn of the two for 13 capture subpopulations and 8 detection targets. Only CoEn meeting Benjamini-Hochberg p-value cutoff (p ≤ 0.01) were included. **(G)** Seven capture-defined EV subpopulations with significant negative dCoEn for Syntenin-1 and ALIX. Error bar represents the progogated variance of each dCoEn.

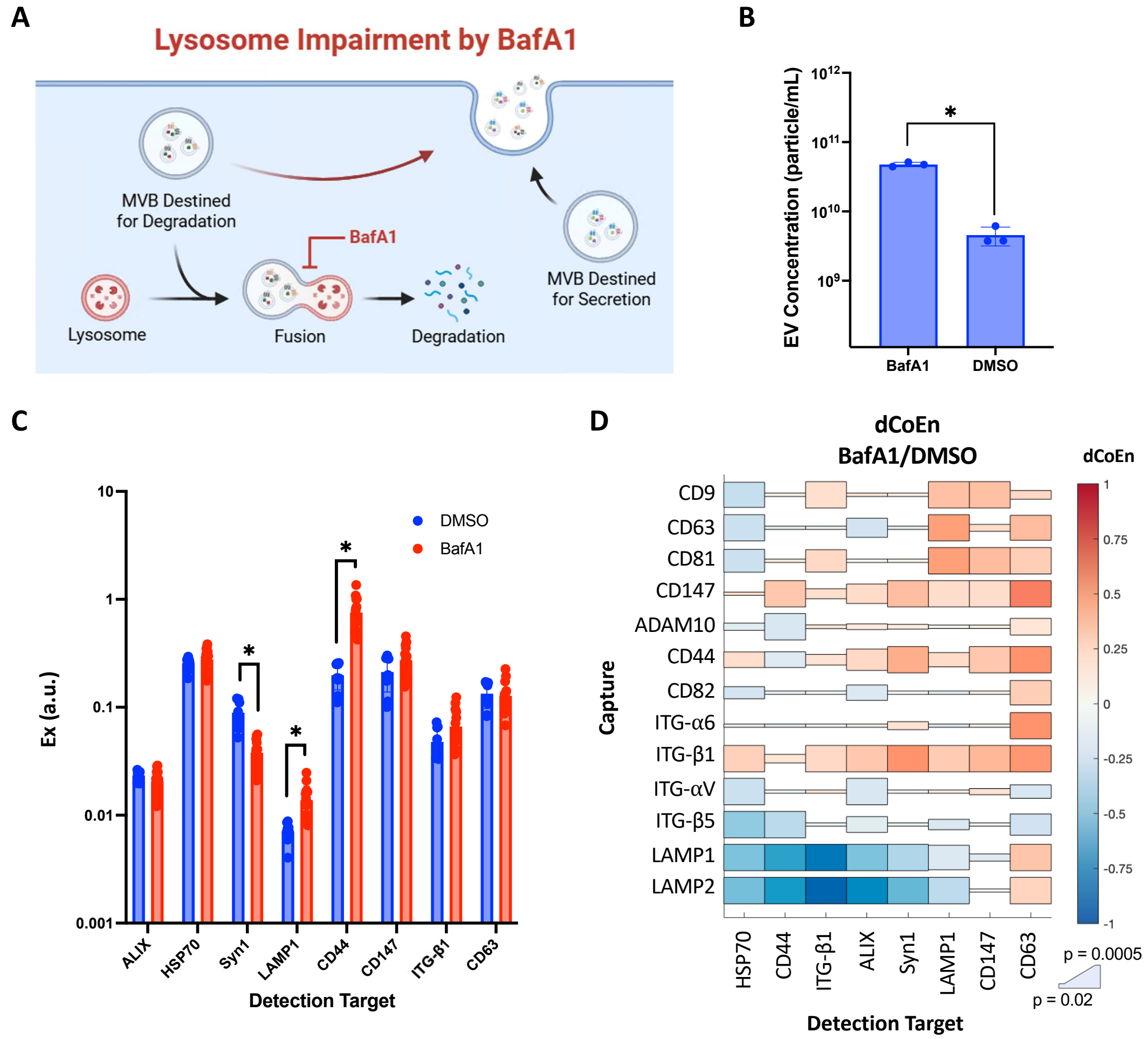

**Figure 5. dCoEn analysis of EVs from 293T cells with and without BafA1 treatment reveals expected and unpredicted changes**. 293T cells were treated with 100 nM of BafA1 or equivalent volume of DMSO for 48 h. **(A)** Schematic of of BafA1-induced EV biogenesis alteration. Upon BafA1 treatment, EVs destined for lysosomal degradation are secreted which is expected to alter protein CoEn in secreted EVs. **(B)** Particle concentration measured by Nanoparticle tracking analysis (NTA) of BafA1-treated (100 nM) and DMSO-treated 293T EVs after 48h of treatment. N = 3, n = 3 for each conditions. Equal number of 293T cells were treated for each condition in each biological repeat. $p < 0.0001$. **(C)** Ex of all detection targets. (N = 3, n = 5 for the BafA1 treated condition, N = 2, n = 5 for DMSO treated condition). p-Value cutoff ($p = 0.0005$) was determined using the Bonferroni correction. **(D)** heatmap of all 104 dCoEn of EVs of BafA1-treated compared to DMSO-treated cells. Lower significance cutoff of -Log(p-Value) = 1.63 ($p = 0.02$) and upper significance cut-off -Log(p-Value) = 3.31 ($p = 0.0005$) defined based on Benjamini-Hochberg and Bonferroni correction procedures, respectively (N = 3, n = 5 for BafA1 treated condition, N = 2, n = 5 for DMSO treated condition). As expected, CoEn of LAMP1 significantly increased in several EV subpopulations (CD9, CD63, CD81, CD147, CD44 and ITGβ1). Meanwhile, numerous instances of cargo re-routing, as visualized by dCoEn, were observed regardless of the changes in the protein's Ex.

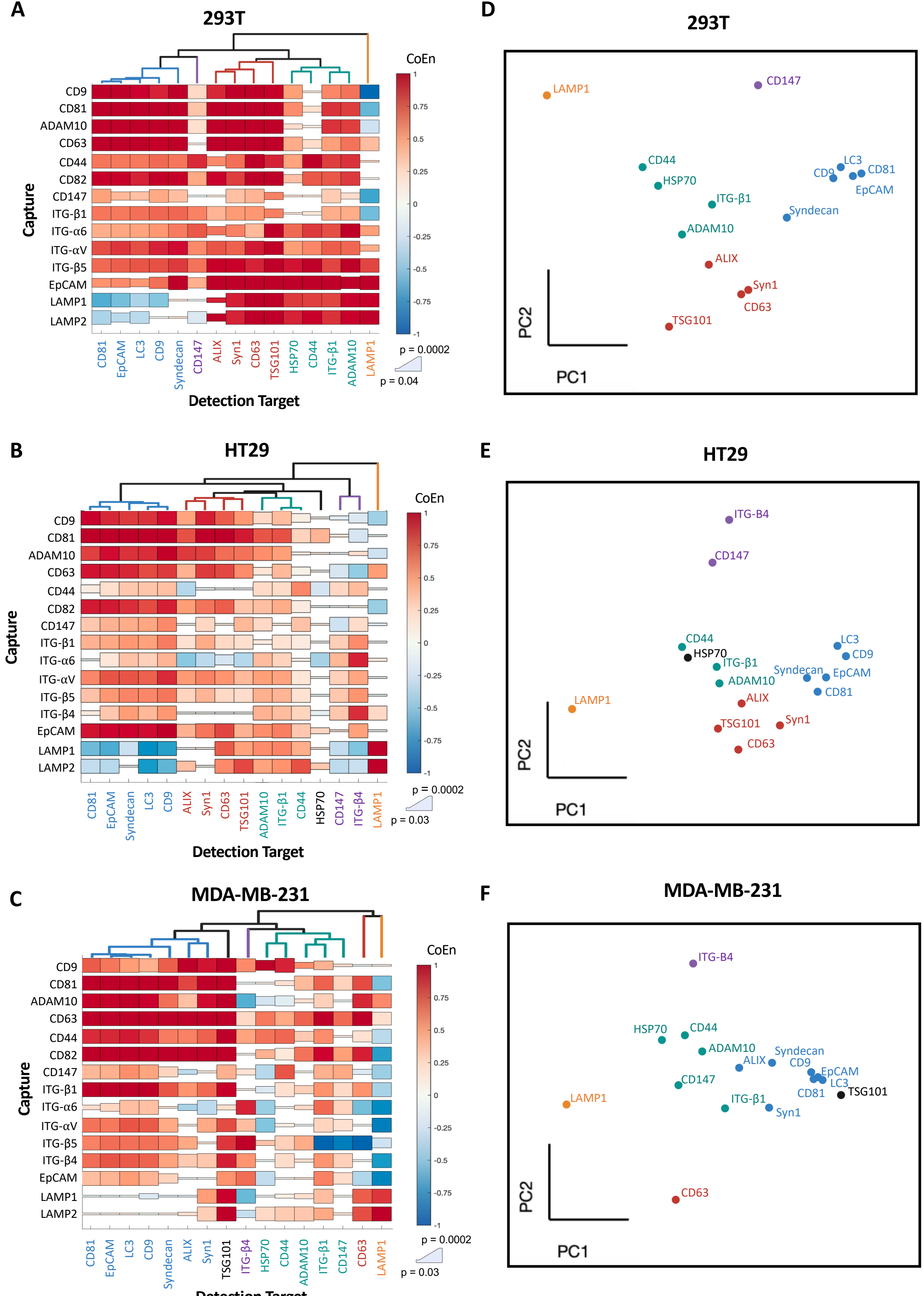

**Figure 6. Hierarchical clustering and PCA analysis of 240 pairwise CoEns of EVs from normal kidney 293T, colorectal cancer HT29, and breast cancer MDA-MB-231 cells identifies conserved groups of proteins**. **(A-C)** Hierarchical clustering based on Euclidian distance of CoEn and heatmap of EVs from 293T **(A)**, HT29 **(B)**, and MDA-MB-231 **(C)**. Note that ITG-B4 is not measured in 293T EVs due to low expression. Lower significance cut-off and upper significance cut-off were based on Benjamini-Hochberg and Bonferroni correction, respectively. N = 3, n = 5 for all three cell line EVs. **(D-F)** Principal component analysis (PCA) based on the pan-capture CoEn distance of each detection target. Hierarchical clustering and PCA identify 5 main groups that are distinguished by putative association of proteins with different pathways. Referring to **E**, Group 1 (LAMP1) is presumptively lysosomal, Group 2 (Alix, Syn1, CD63 TSG101) is presumptively exosomal, and Group 3 (CD9, CD81, LC3, EpCAM and Syndecan), Group 4 (HSP70, ADAM10, CD44 and ITG-β1) and Group 5 (ITG-β4 and CD147) are presumptively ectosomal/membrane-budded.

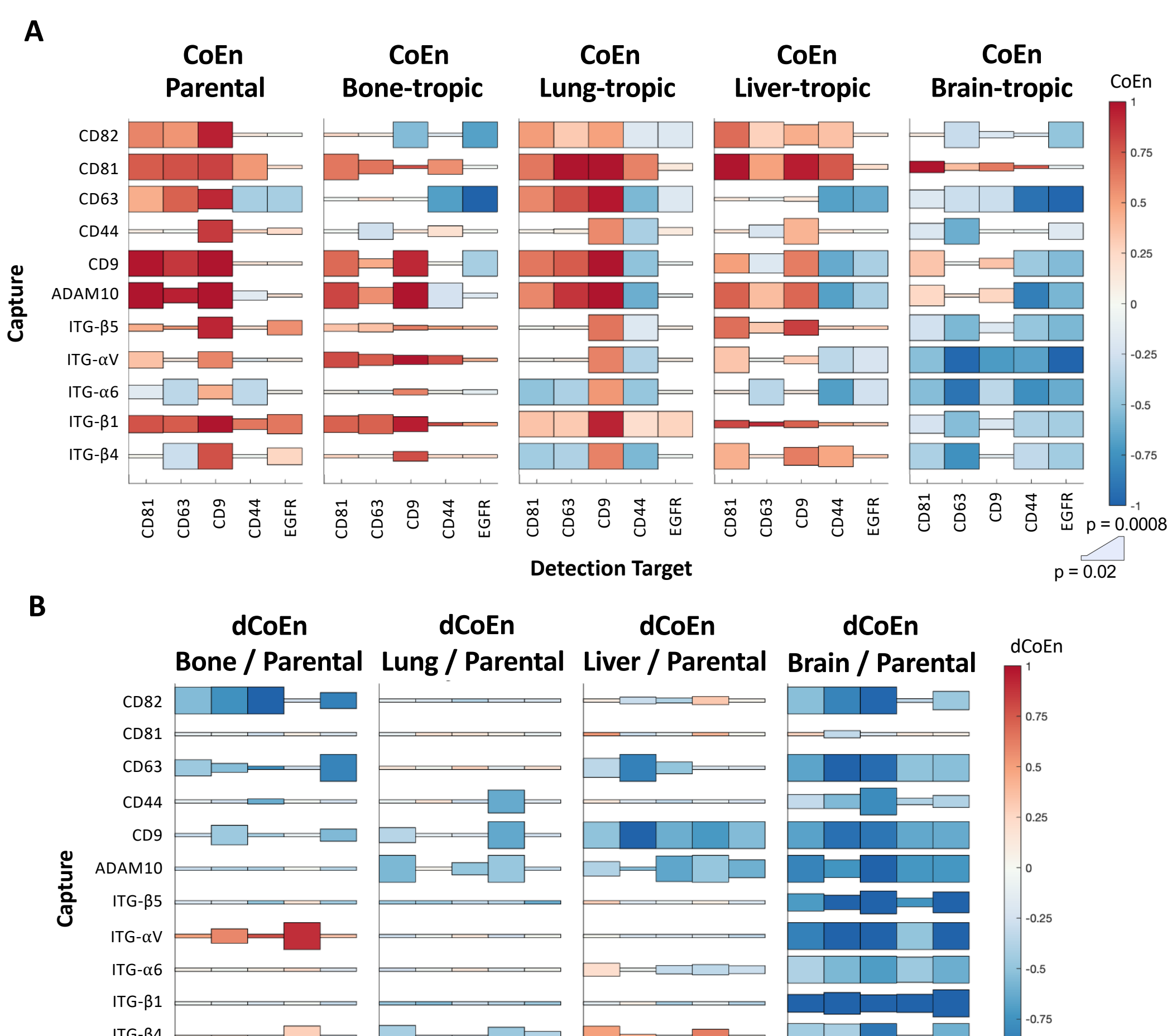

**Figure 7. CoEn and dCoEn of proteins in EVs from a parental breast cancer cell line and organotropic sublines reveal tropism-specific differences.** The parental line is the triple negative breast cancer MDA-MB-231 and organotropic sublines isolated by artificial selection include bone- (1833), lung- (4175), liver- (6113) and brain-tropic (831). **(A)** CoEn and **(B)** dCoEn heatmap of EVs from parental and organotropic sublines. Lower significance cutoff of -Log(p-Value) = 1.64 (p = 0.02) and upper significance cut-off -Log(p-Value) = 3.07 (p = 0.0008) defined based on Benjamini-Hochberg and Bonferroni correction, respectively. N = 3, n = 5 for all but bone-tropism, N = 2, n = 5 for bone tropism (1 biological repeat removed due to microarray functionalization abnormality). CoEn encompass both positive and negative values, while dCoEn are predominantly negative with few notably exceptions including CD44, ITGαV and CD63. Brain organotropic cells stand out by their strong bias towards negative CoEn and dCoEn values.

# Co-Enrichment of Proteins in Extracellular Vesicles

Molly L. Shen[1-4], Andreas Wallucks[1,2], Rosalie Martel[1,2,†], Zijie Jin[1,2,†], Lucile Alexandre[1,2], Philippe DeCorwin-Martin[1,2], Lorenna Oliveira Fernandes de Araujo[1,2], Andy Ng[1,2], Peter M. Siegel[3,4], and David Juncker[1,2,*]

[1]Biomedical Engineering Department, McGill University, Montréal, QC, Canada

[2]Victor Phillip Dahdaleh Institute of Genomic Medicine, McGill University, Montréal, QC, Canada

[3]Goodman Cancer Institute, McGill University, Montréal, QC, Canada

[4]Department of Medicine, McGill University, Montréal, QC, Canada

[†] R.M. and Z.J. contributed equally.

## Supplementary Information

**Table of Content:**

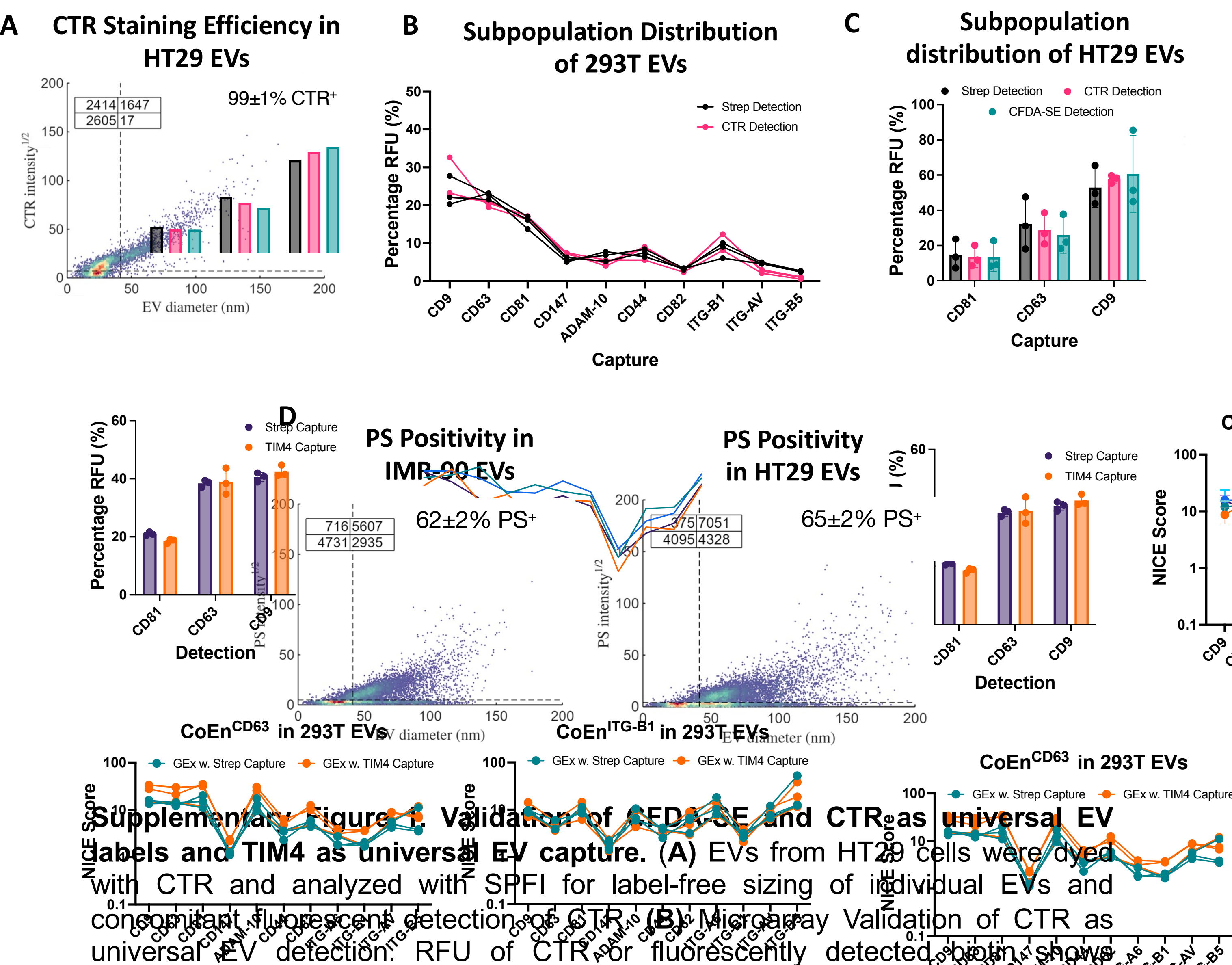

**Supplementary Figure 1. Validation of CFDA-SE and CTR as universal EV labels and TIM4 as universal EV capture.** **(A)** EVs from HT29 cells were dyed with CTR and analyzed with SPFI for label-free sizing of individual EVs and concomitant fluorescent detection of CTR. **(B)** Microarray Validation of CTR as universal EV detection. RFU of CTR for fluorescently detected biotin shows consistent percentage ratio among the 10 EV subpopulations examined. Data shows 3 biological replicates with 5 technical repeats each ( N = 3, n = 5 averaged to 1 value per biological repeat). **(C)** RFU ratio for streptavidin, CTR and CFDA-SE fluorescence for CD81+, CD63+, CD9+ biotinylated EV subpopulation of HT29 EVs. Data shows 3 biological replicates with 10 technical repeats each (biological repeat N = 3, technical repeat n = 10 averaged to 1 value per biological repeats. **(D)** Quantification of Phosphatidylserine (PS) of IMR-90 and HT29 cell line EVs on SPFI by Annexin-FITC staining, revealing that over 62% of EVs are PS+ in each of the cell lines. Scatterplot shows 1 biological repeat with 12 technical repeats (~10,000 single EVs analyzed per cell line).

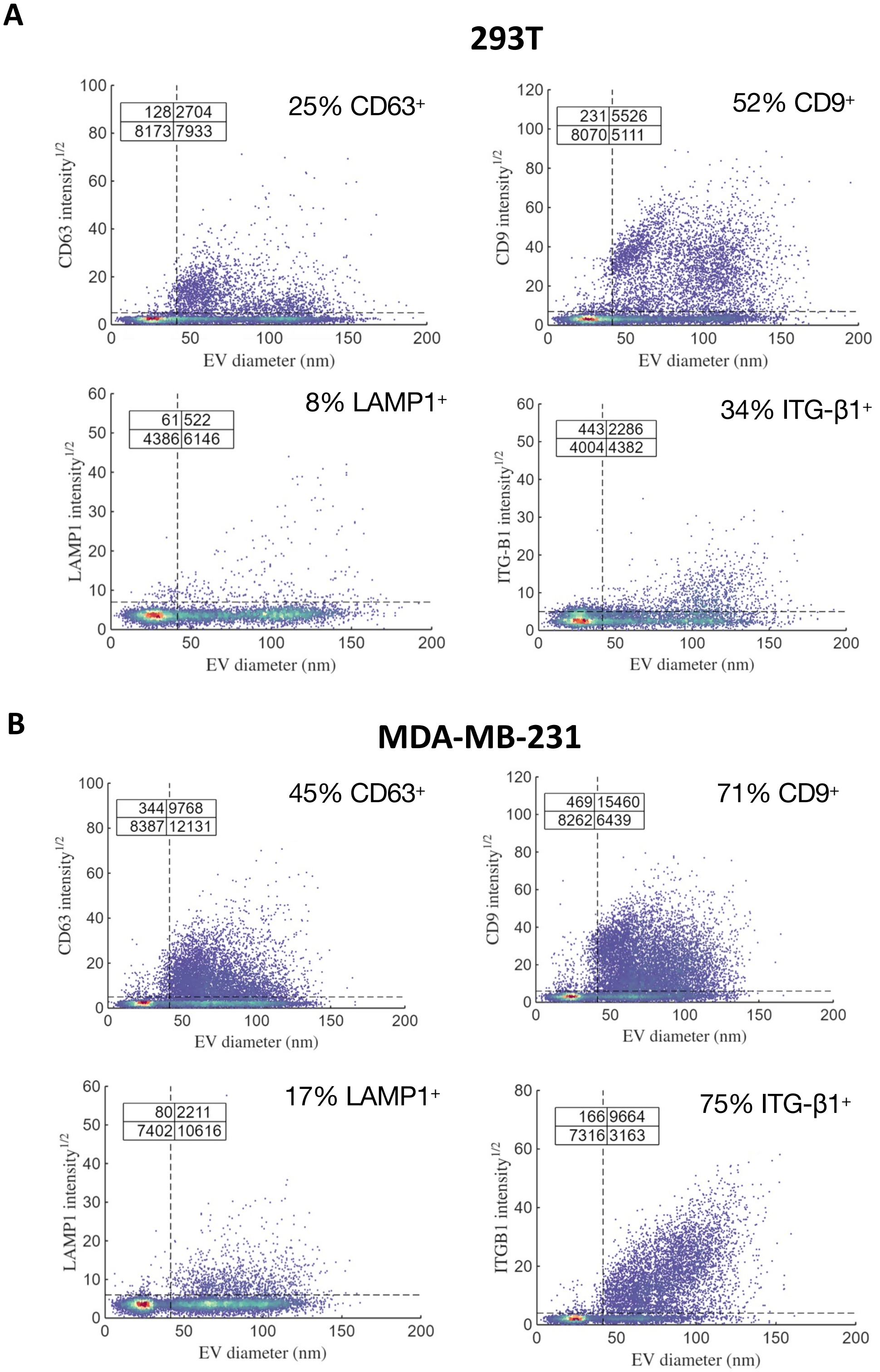

**Supplementary Figure 2. SPFI scatter plots showcase distinct abundance and size preference of the CD63$^+$, LAMP1$^+$, CD9$^+$ and ITG-β1$^+$ EV subpopulations. (A-B)** 293T **(A)** and MDA-MB-231 **(B)** cell lines EVs were immobilized via physisorption on a glass coverslip, sized via SPFI and stained with fluorescent antibodies. Scatterplot shows 1 biological repeat with 7 to 12 technical repeats (~10,000 single EVs analyzed per cell line).

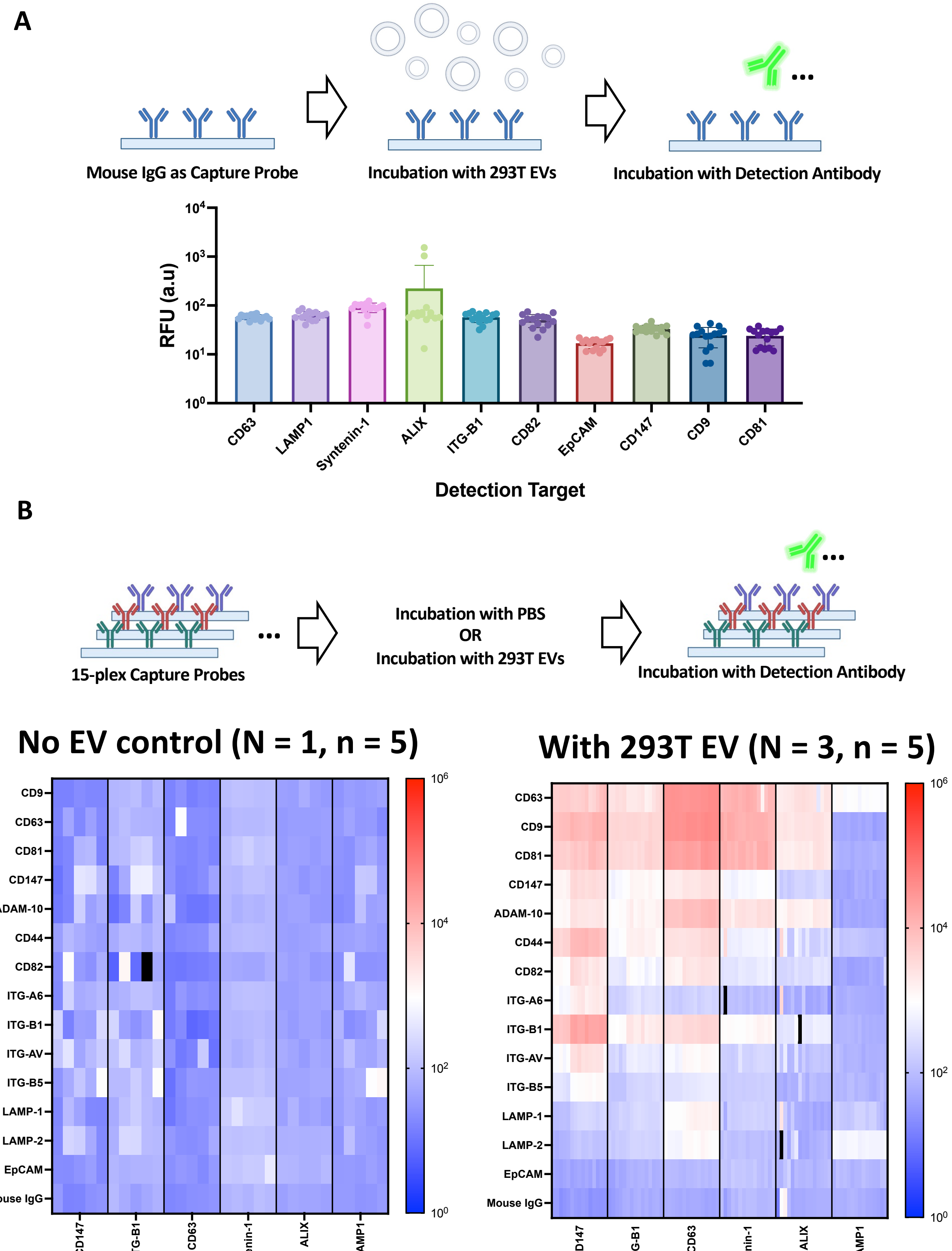

**Supplementary Figure 3. Capture and Detection Antibody Validation. (A)** Establishment of baseline non-specific binding from each detection antibodies using a mouse IgG isotype control as capture probe. **(B)** Evaluation of reagent-driven cross-reactivity between our capture and detection antibodies. Data reported as RFU and plotted in log10 scale. Black boxes indicate zero values which are incompatible with log scaling. Overall, the detection antibodies exhibited comparable levels of non-specific binding to the capture antibodies as observed with mouse IgG, indicating minimal reagent cross-reactivity.

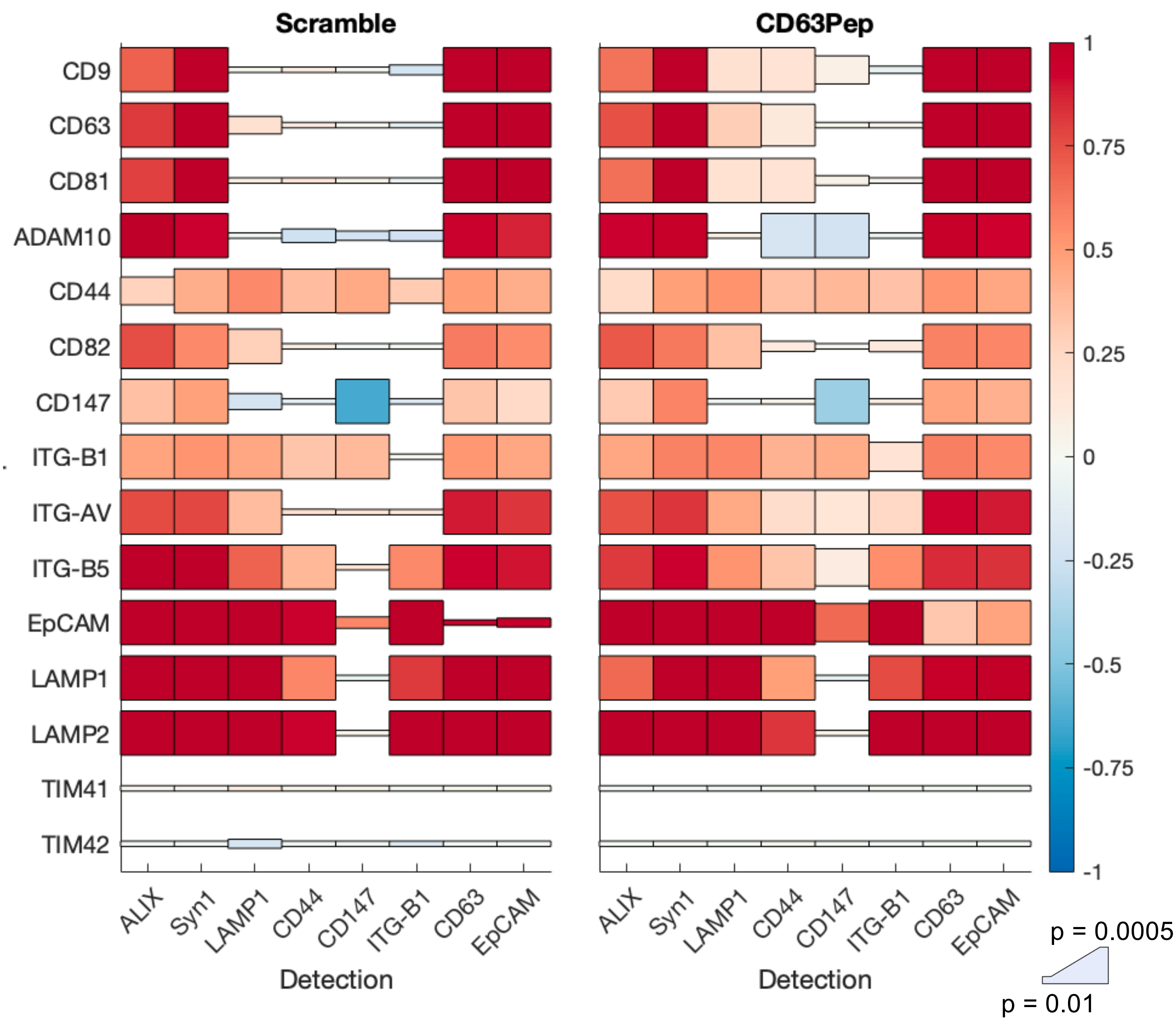

**Supplementary Figure 4. CoEn heatmap of the CD63-peptide-treated and scramble-peptide-treated EV conditions.** Lower significance cutoff of -Log(p-Value) = 1.97 (p = 0.01) and upper significance cut-off -Log(p-Value) = 3.29 (p = 0.0005) defined based on Benjamini-Hochberg and Bonferroni correction procedures. N = 3, n = 5 for CD63 peptide treated condition, N = 2, n = 5 for scrambled peptide treated condition. The two CoEn heatmaps look largely similar owing to the nature of the peptide-induced rerouting. Changes in CoEn between the two conditions is visualized in the dCoEn heatmap shown in Figure 4.

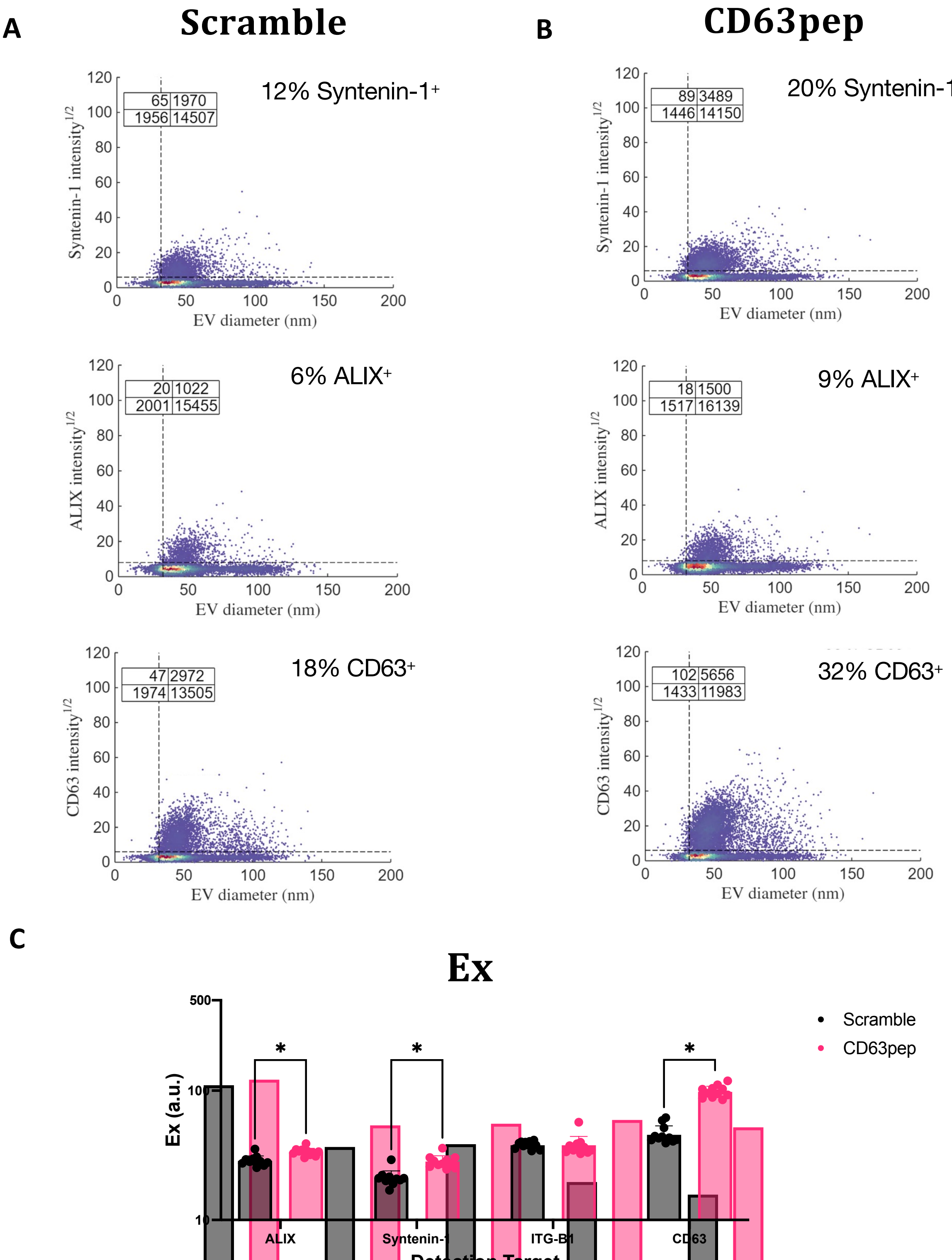

**Supplementary Figure 5. Single EV Quantification of Syntenin-1, ALIX and CD63 positivity and concentration using SPFI. (A-B)** SPFI scatter plots of Syntenin-1, ALIX and CD63 expression in **(A)** scramble-peptide-treated and **(B)** CD63-peptide-treated 293T EVs. EVs were immobilized via physisorption on a poly-L-lysine-coated coverslip, sized via SPFI and stained with fluorescent antibodies. Scatterplot shows 1 biological repeat with 12 technical repeats (~20,000 single EVs analyzed per condition). **(C)** Global expression, Ex, derived by dividing the total RFU signal of marker positive EVs by the number of total EVs analyzed. Asterisk indicates $p < 0.0005$.

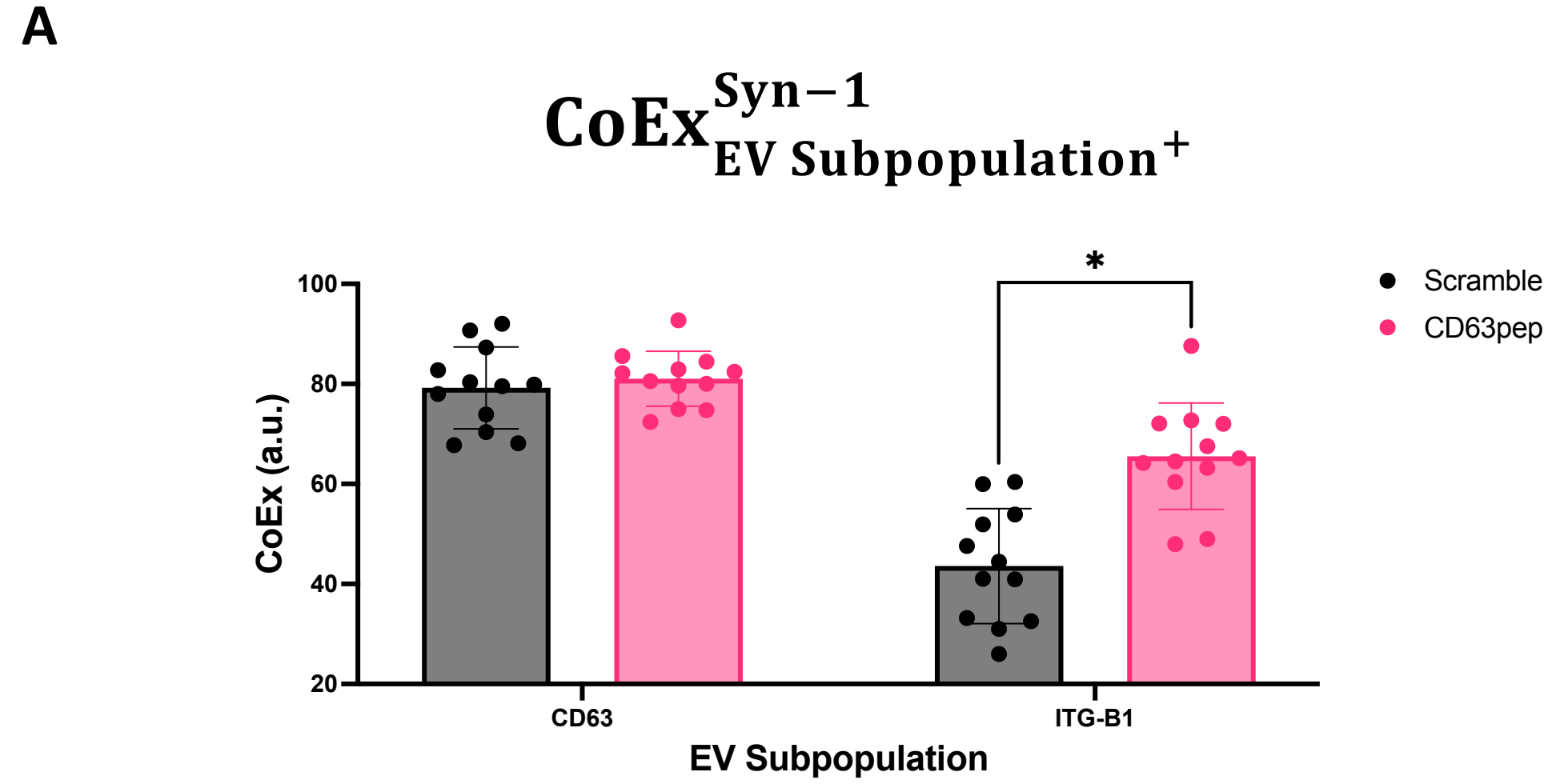

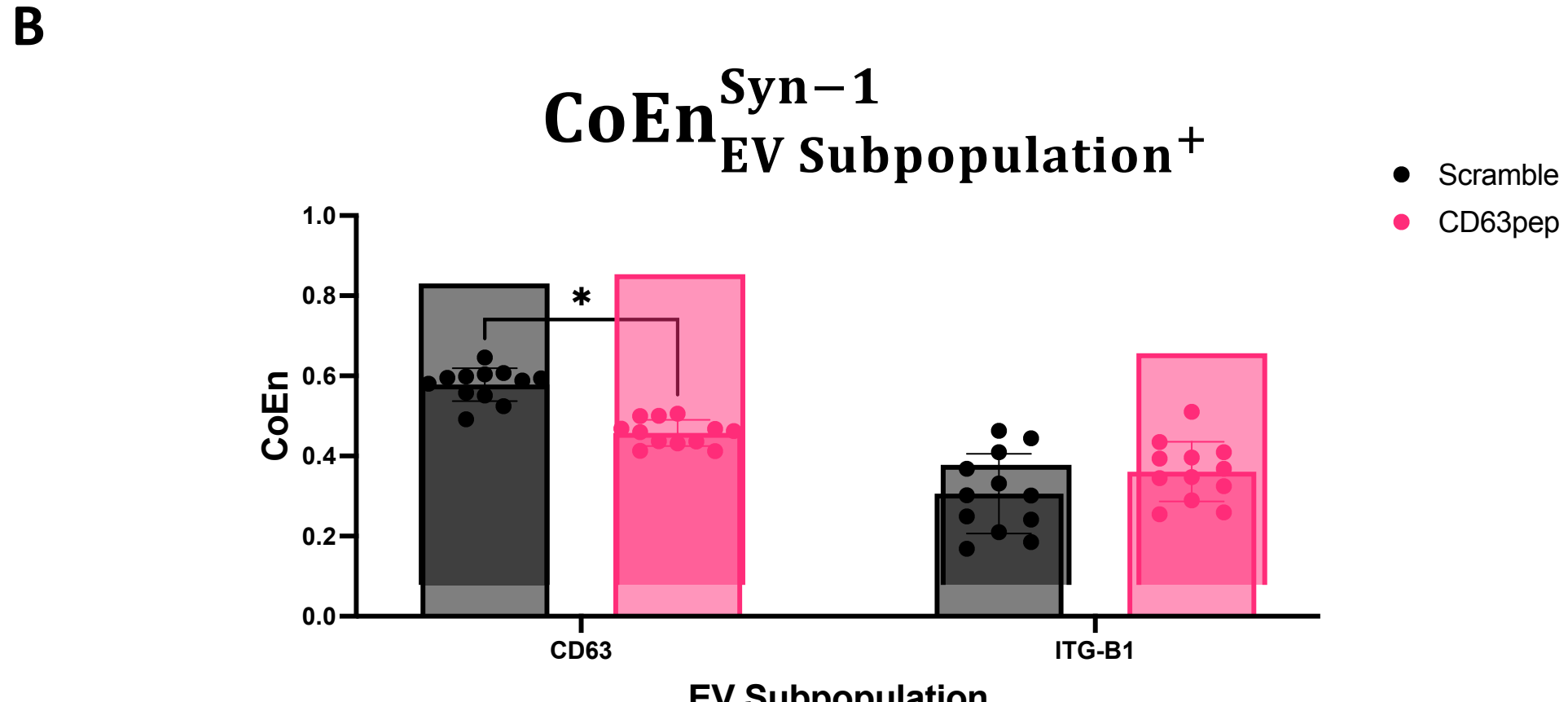

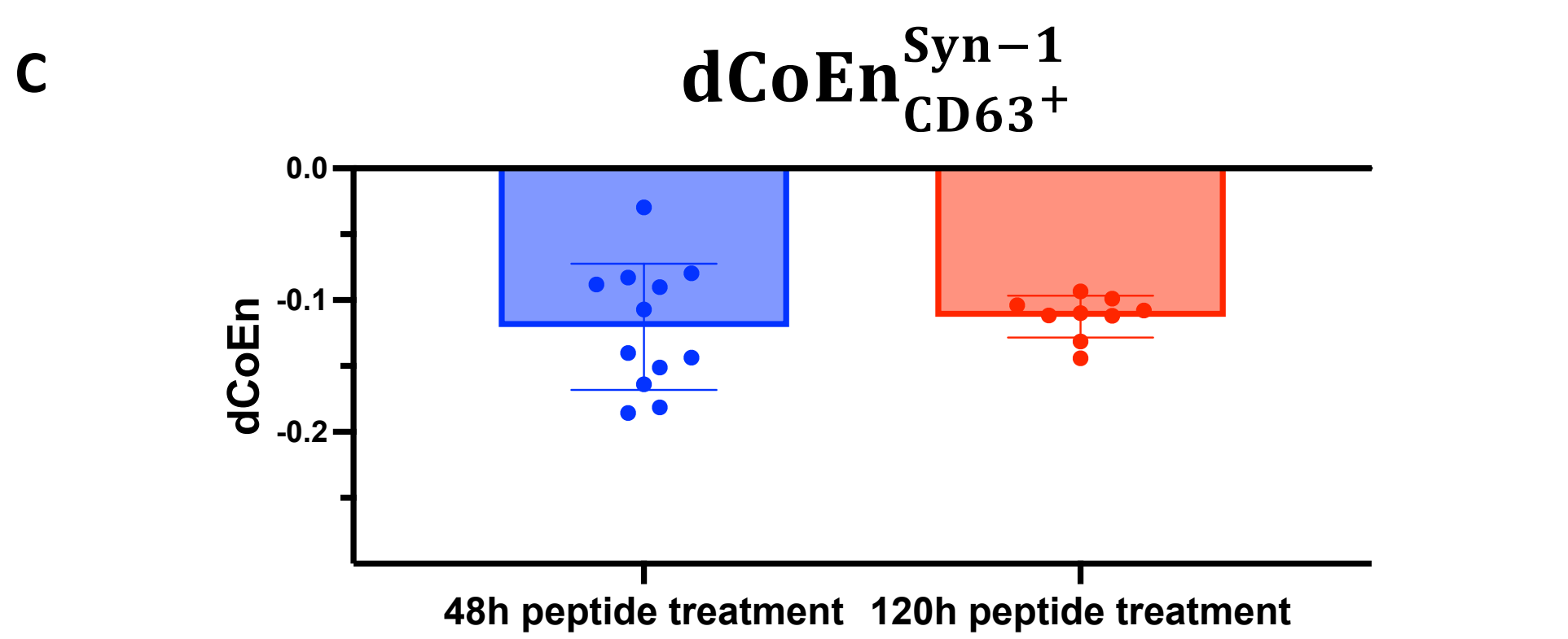

**Supplementary Figure 6. Single EV validation of microarray-visualized Syntenin-1 dCoEn shift via a peptide-induced blockade of Syntenin-CD63 binding**. **(A)** Single-EV-derived CoEx and **(B)** CoEn of Syntenin-1 in CD63$^+$ and ITG-β1$^+$ EVs in CD63- or scrambled-peptide treated 293T. $\mathrm{CoEx}^{\mathrm{A}}_{\mathrm{B}^+}$ is derived by dividing the total RFU signal of A$^+$ EVs by the number of B$^+$ EVs analyzed. Asterisk indicates $p < 0.0005$. **(C)** dCoEn of Syntenin-1 in CD63$^+$ EVs at 48 h and 120 h peptide treatment. ~20,000 single EVs were analyzed from 1 biological repeat with 12 technical repeats.

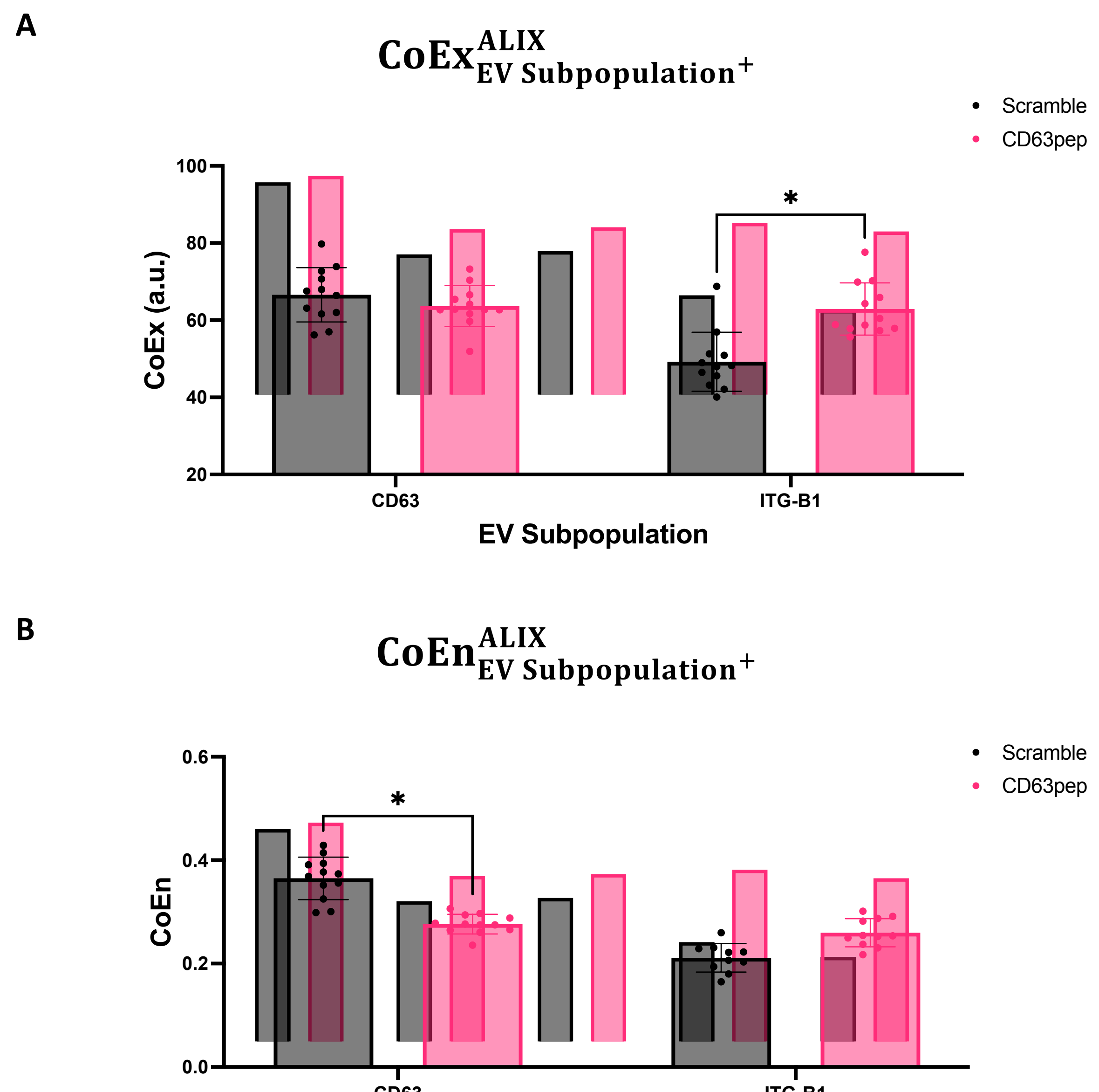

**Supplementary Figure 7. Single EV validation of microarray-visualized ALIX dCoEn shift via a peptide-induced blockade of Syntenin-CD63 binding**. **A)** Single-EV-derived CoEx and **(B)** CoEn of ALIX in CD63$^+$ and ITG-β1$^+$ EVs in CD63- or scrambled-peptide treated 293T. $\mathrm{CoEx}^{\mathrm{A}}_{\mathrm{B}^+}$ is derived by dividing the total RFU signal of A$^+$ EVs by the number of B$^+$ EVs analyzed. Asterisk indicates $p < 0.0005$.

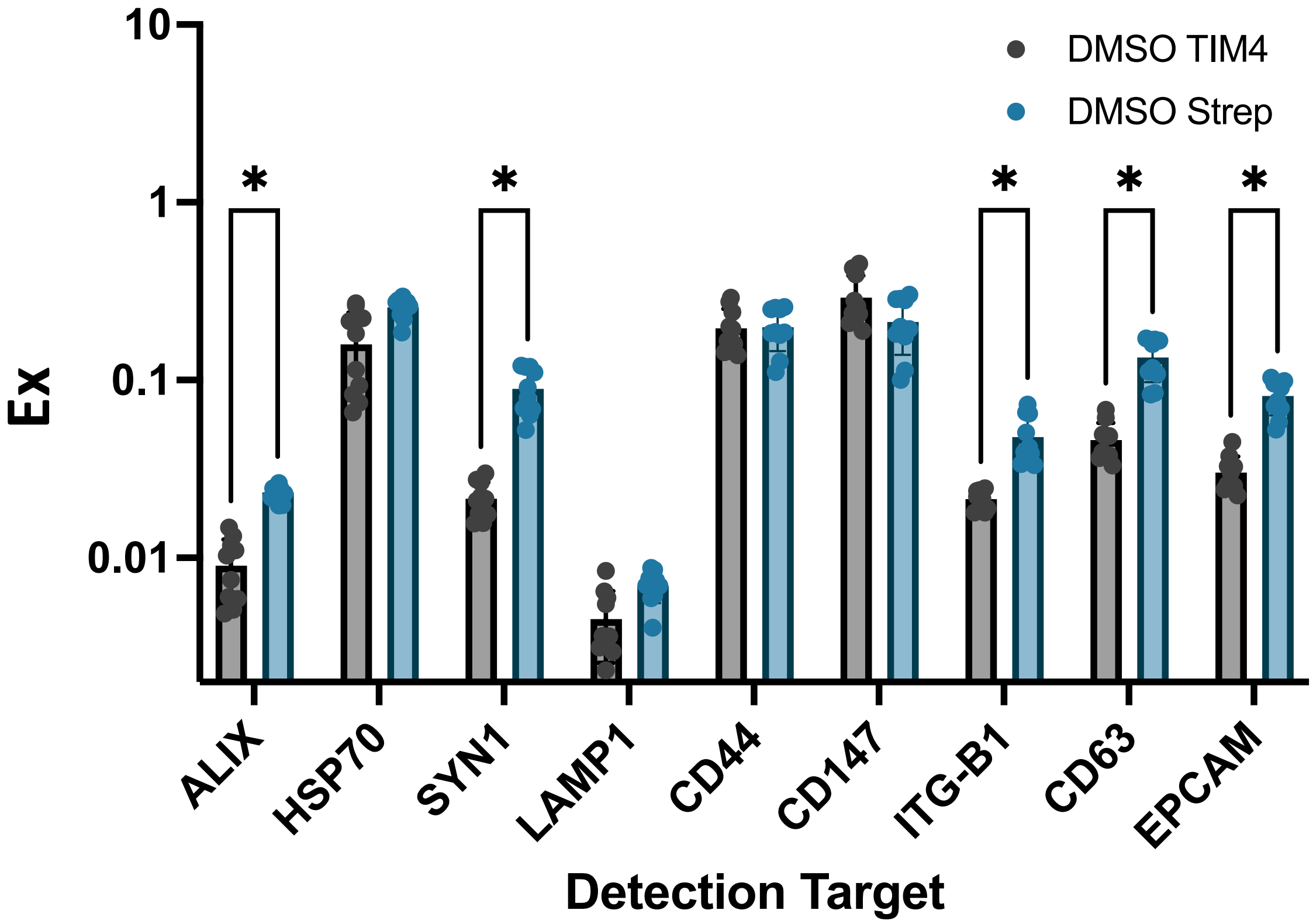

**Supplementary Figure 8. DMSO treatment negate the universality of TIM4 as a pan-EV capture.** Ex generated from TIM4- and streptavidin-capture of biotinylated EVs harvested from DMSO-treated 293T (shown in Figure 5). $p < 0.0005$ (Bonferroni p-value cutoff). As a result of DMSO's effect on TIM4's pan-EV capture capability, EVs were biotinylated and captured with streptavidin in this set of experiment.

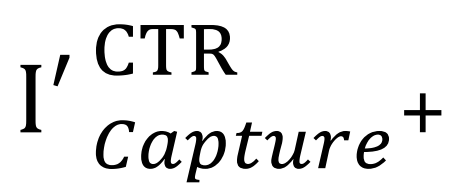

|  | BafA1 | | | | | | | | | | | | | | | DMSO | | | | | | | | | |
|---|---|---|---|---|---|---|---|---|---|---|---|---|---|---|---|---|---|---|---|---|---|---|---|---|---|
| CD9 | 4.37 | 4.232 | 4.227 | 4.143 | 4.277 | 4.055 | 4.098 | 4.06 | 4.073 | 3.894 | 4.107 | 4.088 | 4.046 | 3.986 | 4.29 | 4.313 | 4.309 | 4.311 | 4.32 | 4.295 | 4.224 | 4.265 | 4.235 | 4.271 | 4.238 |
| CD63 | 4.453 | 4.46 | 4.375 | 4.373 | 4.401 | 4.163 | 4.173 | 4.306 | 4.276 | 4.294 | 4.37 | 4.408 | 4.417 | 4.336 | 4.366 | 4.243 | 4.24 | 4.262 | 4.24 | 4.249 | 4.223 | 4.221 | 4.18 | 4.149 | 4.143 |
| CD81 | 4.298 | 4.135 | 4.116 | 4.094 | 4.081 | 4.055 | 4.006 | 3.95 | 3.91 | 3.853 | 3.908 | 3.89 | 4.093 | 4.098 | 4.024 | 4.157 | 4.13 | 4.141 | 4.231 | 4.181 | 4.073 | 4.093 | 4.104 | 4.152 | 4.143 |
| CD147 | 4.183 | 4.063 | 4 | 4.011 | 4.023 | 3.906 | 3.875 | 3.919 | 3.892 | 3.782 | 3.941 | 4.4 | 3.836 | 3.852 | 3.918 | 3.872 | 3.823 | 3.826 | 3.838 | 3.867 | 3.73 | 3.792 | 3.754 | 3.801 | 3.856 |
| LAMP1 | 4.092 | 4.12 | 4.068 | 4.087 | 4.037 | 3.859 | 3.816 | 3.829 | 3.952 | 3.77 | 3.952 | 3.96 | 3.973 | 3.989 | 4.016 | 2.755 | 2.713 | 2.699 | 2.661 | 2.644 | 2.655 | 2.724 | 2.668 | 2.583 | 2.62 |
| ADAM10 | 3.686 | 3.747 | 3.77 | 3.795 | 3.784 | 3.63 | 3.684 | 3.668 | 3.689 | 3.612 | 3.422 | 3.441 | 3.501 | 3.511 | 3.522 | 3.717 | 3.666 | 3.673 | 3.689 | 3.722 | 3.489 | 3.461 | 3.516 | 3.537 | 3.61 |
| CD44 | 4.333 | 4.348 | 4.334 | 4.288 | 4.317 | 4.134 | 4.109 | 4.094 | 4.071 | 4.073 | 4.2 | 4.211 | 4.233 | 4.248 | 4.275 | 3.963 | 3.937 | 3.944 | 3.945 | 3.911 | 3.848 | 3.836 | 3.862 | 3.877 | 3.817 |
| CD82 | 4.064 | 3.762 | 3.819 | 3.902 | 3.822 | 3.768 | 3.694 | 3.807 | 3.782 | 3.696 | 3.434 | 3.393 | 3.524 | 3.588 | 3.59 | 3.473 | 3.472 | 3.473 | 3.473 | 3.519 | 3.319 | 3.368 | 3.37 | 3.384 | 3.356 |
| ITG-A6 | 3.541 | 3.569 | 3.535 | 3.562 | 3.454 | 3.434 | 3.429 | 3.392 | 3.275 | 3.214 | 3.203 | 3.235 | 3.269 | 3.273 | 3.123 | 2.94 | 2.963 | 3.001 | 3.019 | 2.909 | 2.891 | 2.95 | 2.931 | 3.1 | 2.976 |
| ITG-B1 | 4.354 | 4.205 | 4.222 | 4.197 | 4.209 | 4.127 | 4.05 | 4.124 | 4.031 | 4.013 | 3.953 | 3.946 | 3.939 | 3.984 | 3.995 | 4.069 | 4.045 | 4.067 | 4.067 | 4.053 | 4.014 | 4.006 | 3.994 | 3.982 | 4.039 |
| ITG-AV | 3.888 | 3.892 | 3.714 | 3.74 | 3.769 | 3.566 | 3.625 | 3.668 | 3.606 | 3.522 | 3.421 | 3.435 | 3.455 | 3.504 | 3.501 | 3.506 | 3.462 | 3.439 | 3.477 | 3.459 | 3.368 | 3.412 | 3.371 | 3.325 | 3.348 |
| ITG-B5 | 3.385 | 3.371 | 3.426 | 3.367 | 3.376 | 3.164 | 3.193 | 3.171 | 3.53 | 3.058 | 3.078 | 3.043 | 3.57 | 3.787 | 3.026 | 3.092 | 3.052 | 3.04 | 3.075 | 3.042 | 2.907 | 2.921 | 2.982 | 2.946 | 2.865 |
| LAMP2 | 3.887 | 3.889 | 3.927 | 3.948 | 3.979 | 3.804 | 3.873 | 3.854 | 3.834 | 3.824 | 3.698 | 3.703 | 3.843 | 3.76 | 3.848 | 2.402 | 2.462 | 2.37 | 2.339 | 2.422 | 2.245 | 2.262 | 2.448 | 2.326 | 2.314 |
| TIM41 | 4.446 | 3.995 | 4.275 | 3.983 | 3.992 | 3.853 | 3.737 | 3.736 | 4.038 | 4.408 | 3.962 | 3.918 | 3.987 | 4.429 | 4.151 | 3.965 | 3.993 | 3.924 | 3.869 | 3.893 | 3.983 | 3.965 | 4.1 | 3.928 | 3.931 |
| TIM42 | 4.328 | 4.007 | 3.943 | 3.912 | 3.994 | 4.474 | 4.193 | 3.787 | 3.856 | 3.809 | 3.994 | 4.284 | 4.09 | 3.975 | 3.935 | 3.917 | 3.952 | 3.907 | 3.909 | 3.942 | 3.888 | 3.951 | 3.945 | 3.948 | 3.992 |

4.5 4 3.5 3 2.5 2 1.5 1

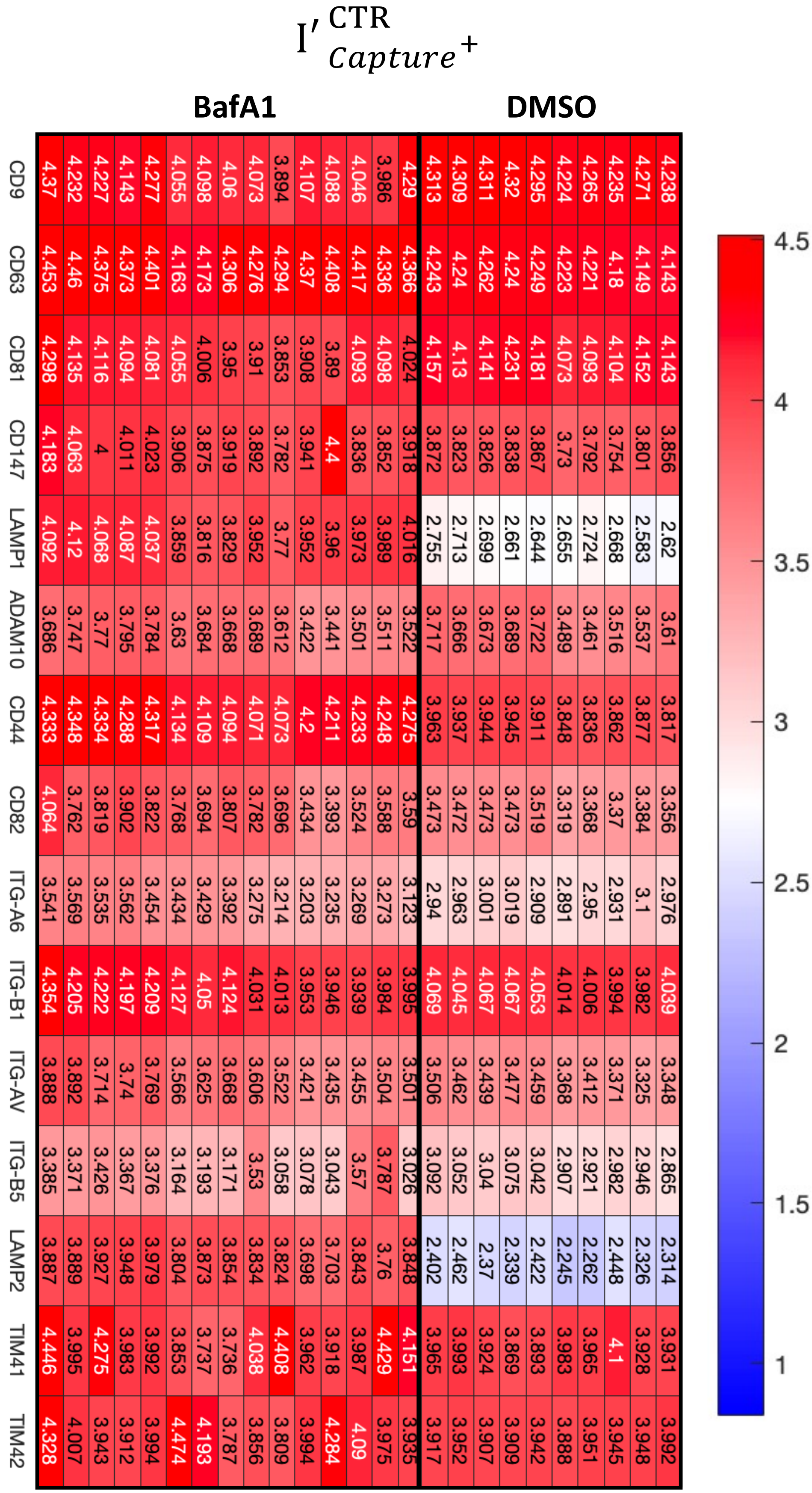

**Supplementary Figure 9. Bafilomycin A1 treatment increased the subpopulation size of LAMP1⁺ and LAMP2⁺ EVs.** Raw CTR RFU of all antibody-captured EV subpopulations in Bafilomycin A1 and DMSO-treated 293T (shown in Figure 5). N = 3, n = 5 for BafA1 treated condition, N = 2, n = 5 for DMSO treated condition.

$$\mathbf{CoEn}_{\mathbf{Capture}}^{Detection}$$

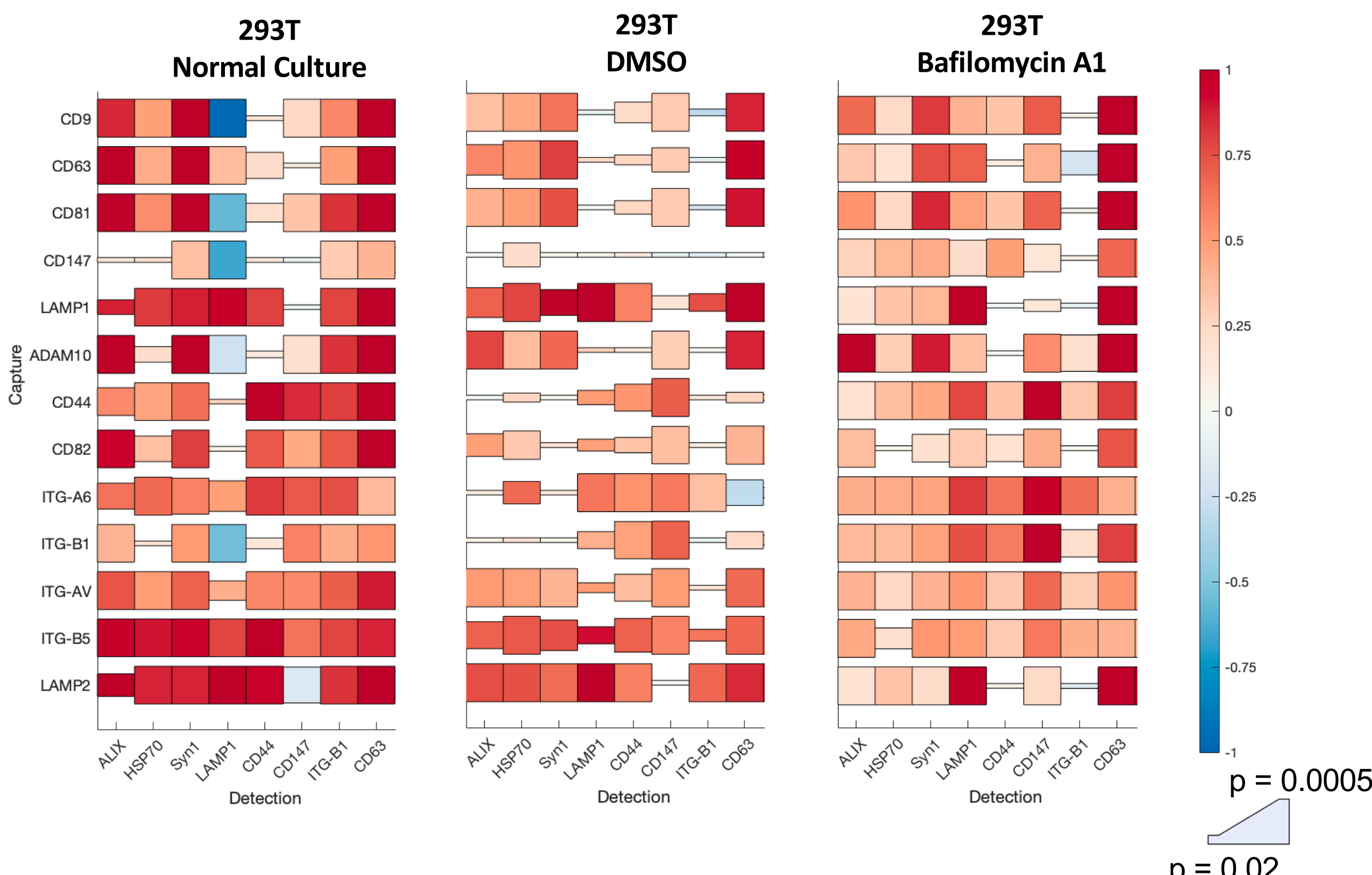

**Supplementary Figure 10. CoEn heatmap of EVs from 293T cells cultured in regular media, with DMSO and with BafA1**. Lower significance cutoff of -Log(p-Value) = 1.63 (p = 0.02) and upper significance cut-off -Log(p-Value) = 3.31 (p = 0.0005) defined based on Benjamini-Hochberg and Bonferroni correction procedures. N = 3, n = 5 for regular media and BafA1 treated condition, N = 2, n = 5 for DMSO treated condition. The CoEn of LAMP1 was somewhat unexpectedly altered by the presence of DMSO, as well as with Bafilomycin A1.

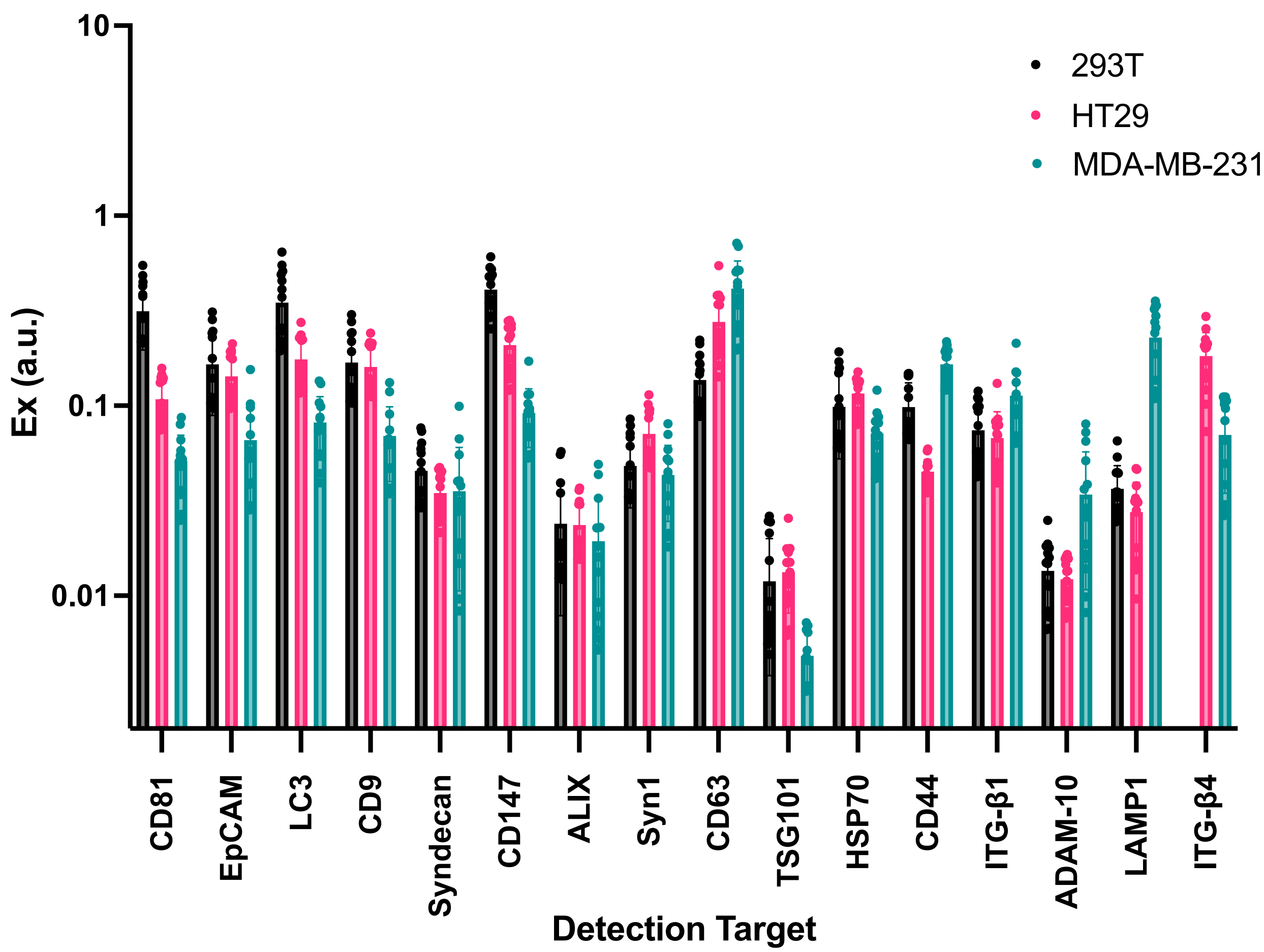

**Supplementary Figure 11. Ex of 16 proteins for EVs detection proteins in 293T EVs and 16 detection targets in HT29 and MDA-MB-231 EVs.** N =3, n = 5, each dot represents a single technical repeat. ITG-B4 was not measured in 293T as it was undetectable. Overall, the EV Ex profiles differed among the three cell lines, with notable variations in protein markers such as LAMP1, CD81, and CD147.

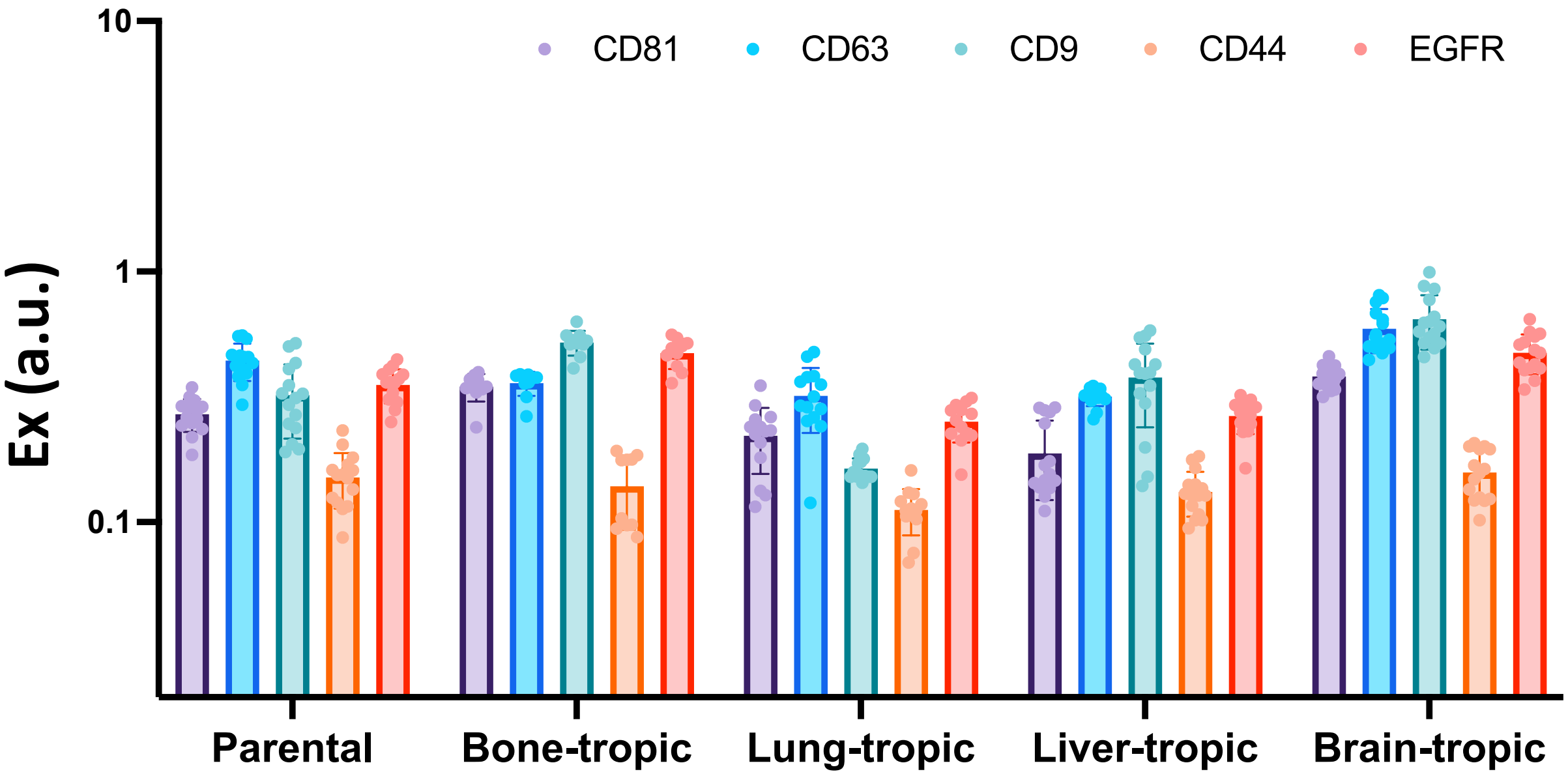

**Supplementary Figure 12. Ex of tetraspanins CD81, CD63, CD9 and cancer associated markers CD44 and EGFR proteins in EVs derived from the triple negative breast cancer cell line MDA-MB-231 (PAR) and its bone- (1833), lung- (4175), liver- (6113) and brain- (831) organotropic sublines.** Each dot represents a single technical repeat (N = 2, n = 5 for bone-tropism; N = 3, n = 5 for all other conditions). The EV Ex profile between parental and its organotropic sublines are relatively similar.

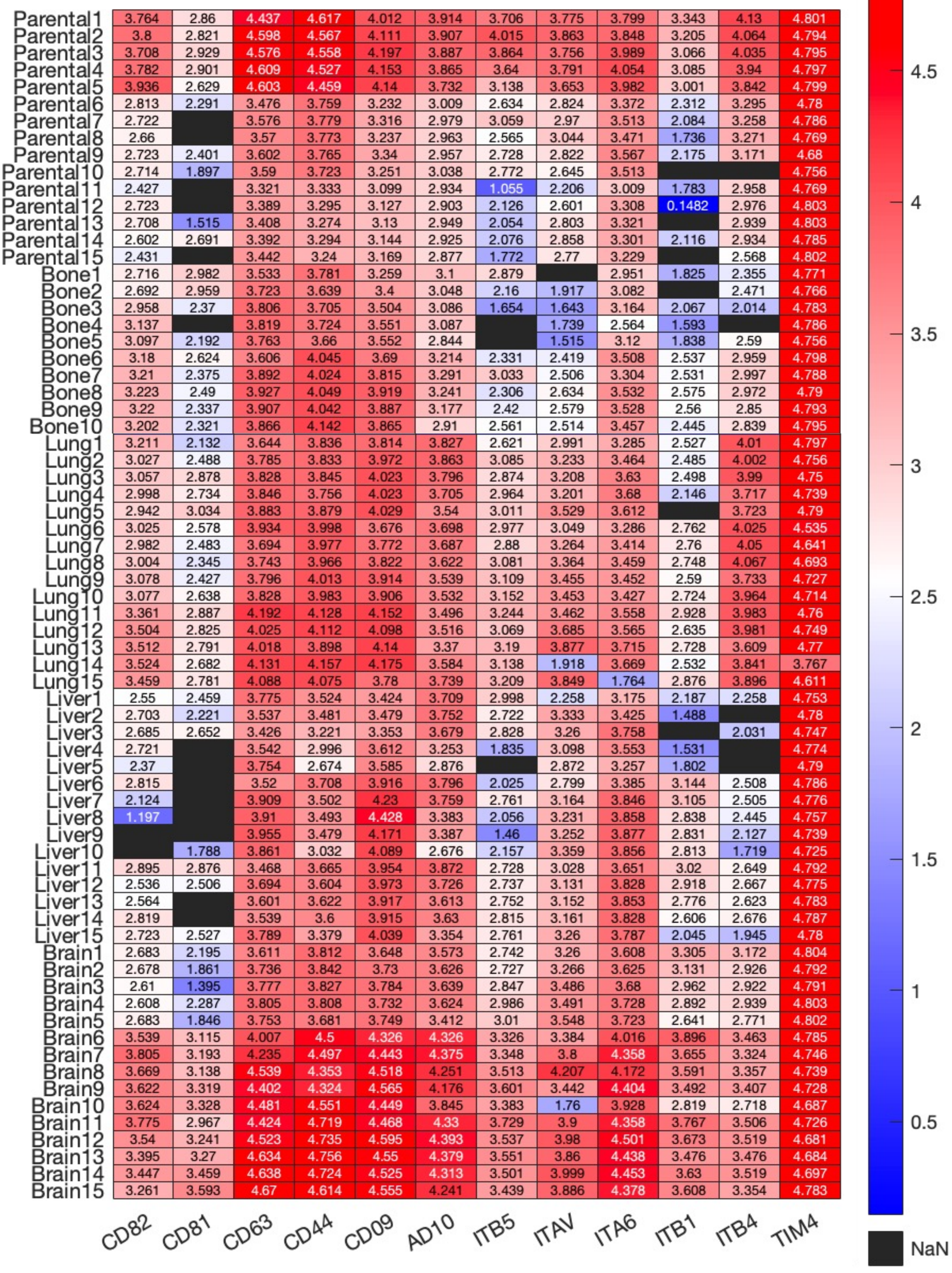

**Supplementary Figure 13. Raw CTR RFU of all antibody-captured EV subpopulations for the organotropic cell line EVs.** One outlier As biological repeat from bone-tropic 1833 was removed due to microarray functionalization abnormality. Lung-tropic 4175 subline EVs had the most abundant ITG-β4 subpopulation.

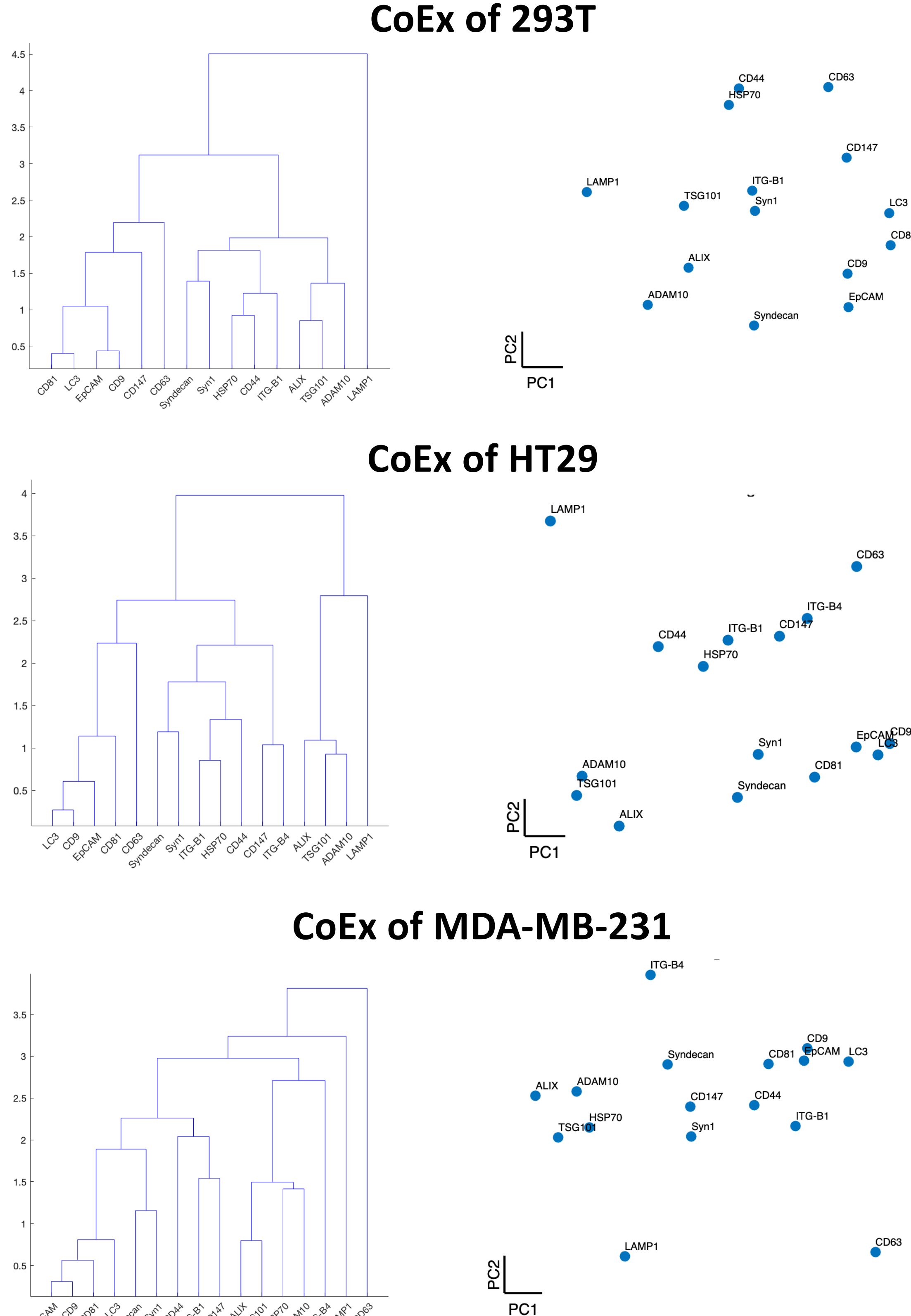

**Supplementary Figure 14. Hierarchical clustering and PCA analysis of CoEx values for EVs of 293T, HT29 and MDA-MB-231.** Hierarchical Clustering was done with Euclidean distance of CoEx scores in $Log_{10}$. Principal component 1 vs. 2 (PC1 vs. PC2) graph of the z-scored pan-capture $Log_{10}$ CoEx patterns of each detection targets. While hierarchical clustering and PCA analysis of CoEn patterns led to conserved grouping of proteins that are consistent with the literature and provide new insights, CoEx patterns failed to visualize these protein groups.

| Target | Type | Species | Supplier | Catalog # | Lot # | Clone |
|---|---|---|---|---|---|---|
| ADAM-10 | Monoclonal | Mouse | R&D Systems | MAB1427 | HZR051710B | 163003 |
| CD44 | Monoclonal | Rat | Biolegend | 103033 | B269039 | IM7 |
| CD44 (Detection) | Monoclonal - Biotin | Rat | Biolegend | 103003 | B187391 | IM7 |
| CD63 | Monoclonal | Mouse | Biolegend | 353014 | B213442 | H5C6 |
| CD63 (Detection) | Monoclonal - Biotin | Mouse | BioLegend | 353018 | B242325 | H5C6 |
| CD81 | Monoclonal | Mouse | Biolegend | 349502 | B264968 | 5A6 |
| CD81 (Detection) | Monoclonal - Biotin | Mouse | Biolegend | 349514 | B232723 | - |
| CD82/Kai-1 | Monoclonal | Mouse | R&D systems | MAB4616 | ZVV061710B | 423524 |
| CD9 | Monoclonal | Mouse | Biolegend | 312102 | B279344 | HI9a |
| CD9 (Detection) | Monoclonal - Biotin | Mouse | Biolegend | 312112 | B266570 | - |
| EGFR | Monoclonal | Mouse | Biolegend | 617502 | B286054 | A18011A |
| EGFR (Detection) | Monoclonal Biotin | Goat | R&D systems | BAF231 | AUD1317081 | - |
| EpCAM (Detection) | Polyclonal - Biotin | Goat | R&D systems | BAF960 | FOZ2119101 | - |
| EpCAM | Polyclonal | Goat | R&D Systems | AF960 | FOX0817101 | - |
| GFP (Capture control) | Polyclonal | Rabbit | Invitrogen | PA5-22688 | SK2470601 | - |
| Integrin α1 | Monoclonal | Mouse | R&D systems | MAB5676 | CDZO0215121 | 639508 |
| Integrin α1 (Detection) | Polyclonal - Biotin | Rabbit | Mybiosource | MBS7104944 | G0612D1 | - |
| Integrin α2 | Monoclonal | Rat | R&D systems | MAB12332 | CBDM0118021 | 430903 |
| Integrin α2 (Detection) | Monoclonal - Biotin | Mouse | R&D systems | BAM1233 | JSZ0115041 | - |
| Integrin α5 | Monoclonal | Mouse | R&D systems | MAB1864 | JZX0319041 | 238307 |
| Integrin α5 (Detection) | Polyclonal - Biotin | Goat | R&D systems | BAF1864 | JZY0217121 | - |
| integrin α6 | Monoclonal | Mouse | R&D Systems | MAB1350 | HNO051710A | MP4F10 |
| integrin α6(Detection) | Monoclonal | Rat | R&D systems | MAB13501 | INVV0118081 | GoH3 |
| Integrin αV | Monoclonal | Mouse | R&D systems | MAB1219 | HJH0218071 | P2W7 |
| Integrin αV (Detection) | Polyclonal - Biotin | Goat | R&D systems | BAF1219 | YOW0317121 | - |
| integrin β1 | Monoclonal | Mouse | R&D Systems | MAB17781 | UUY0316112 | P5D2 |
| Integrin β1 (Detection) | Polyclonal | Goat | R&D systems | AF1778 | WGJ0418111 | - |
| Integrin β3 | Monoclonal | Mouse | R&D systems | MAB2266 | KUT0219071 | 256809 |
| Integrin β3 (Detection) | Polyclonal - Biotin | Goat | R&D systems | BAF2266 | UKJ0317112 | - |
| Integrin β4 | Monoclonal | Mouse | R&D systems | MAB4060 | ZNC0219021 | 422325 |
| Integrin β4 (Detection) | Monoclonal - Biotin | Rat | Abcam | ab95584 | GR3297241-1 | 439-9B |
| Integrin β5 | Monoclonal | Mouse | ThermoFisher | 14-0497-82 | 2043810 | KN52 |
| Integrin β5 (Detection) | Polyclonal - Biotin | Sheep | R&D systems | BAF3824 | YJR0112051 | - |
| Integrin β6 | Monoclonal | Mouse | R&D systems | MAB4155 | ZVW0218121 | 437211 |
| Integrin β6 (Detection) | Polyclonal - Biotin | Sheep | R&D systems | BAF4155 | ZQZ0319041 | - |
| TIM4-Fc Chimera | Recombinant | Human | R&D Systems | 9300-TM | DFYL0520091 | Q96H15 |
| TSG101 (Detection) | Monoclonal-oligo conjugated | Mouse | invitrogen | MA1-23296 | YK4103742A | 4A10 |
| LC3 (Detection) | Monoclonal-oligo conjugated | Mouse | Biolegend | 848802 | B312340 | A15143K |
| Syntenin-1 (Detection) | Monoclonal-oligo conjugated | Mouse | invitrogen | MA5-32932 | YC3852151A | 2B8 |
| Syndecan (Detection) | Monoclonal-oligo conjugated | Mouse | Biolegend | 352302 | - | DL-101 |
| LAMP1 | Monoclonal | Mouse | Biolegend | 328602 | B366026 | H4A3 |
| LAMP2 | Monoclonal | Mouse | Biolegend | 354302 | - | H4B4 |
| LAMP1 (Detection) | Monoclonal-oligo conjugated | Mouse | Biolegend | 328602 | B366026 | H4A3 |
| CD147 | Monoclonal | Mouse | Biolegend | 306202 | B262608 | H1M6 |
| CD147 (Detection) | Monoclonal-oligo conjugated | Mouse | Biolegend | 306202 | B262608 | H1M6 |
| Alix (Detection) | Polyclonal-Oligo conjugated | Rabbit | Invitrogen | PA5-52873 | YF3956674 | AB_2637865 |
| HSP70 (Detection) | Polyclonal-Oligo conjugated | Rabbit | Invitrogen | PA5-34772 | WD3258269B | AB_2552124 |

**Supplementary Table 1. Protein targets and antibodies used in this work.**

**Supplementary Method:**
**Detailed Workflow and Calculation of Ex, CoEx and CoEn using ELISA.**

**Workflow:**

1. **Experimental Derivation of $\mathbf{Ex^A}$ :**
    a. The **Ex** of protein A can be determined using a standard sandwich ELISA targeting protein A on a defined quantity of lysed EVs (e.g., 10 µg of EVs).

*Example readout:* $\mathrm{Ex^A} = 50\ pg/mL$

2. **Experimental Derivation of $\mathbf{CoEx^A_{B^+}}$ :**
    **a. $B^+$ EVs isolation** can be performed using any commercially available superparamagnetic bead system (e.g., Dynabeads):
        a. Functionalize beads with anti-B antibodies using any surface chemistry of your choice (e.g., streptavidin beads & biotinylated antibodies)
        b. Incubate EV samples with anti-B functionalized beads
        c. Elute $B^+$ EVs
    *b.* $\mathbf{CoEx^A_{B^+}}$ can be quantified by performing a sandwich ELISA targeting protein A on a fixed quantity of lysed $B^+$ EVs (e.g., 10 µg of $B^+$ EVs).

*Example readout:* $\mathrm{CoEx^A_{B^+}} = 500\ pg/mL$

*3.* $\mathbf{CoEn^A_{B^+}}$ is then calculated as:

$$\mathrm{CoEn^A_{B^+}} = \log_{10}\left(\frac{\mathrm{CoEx^A_{B^+}}}{\mathrm{Ex^A}}\right)$$

*For our Examples*: $\log_{10}\left(\frac{500\ pg/mL}{50\ pg/mL}\right) = 1$

**Technical Notes:**

1. EVs can be quantified and normalized either by particle count using particle counting instruments (e.g., MRPS, NTA) or by total protein quantification (e.g., microBCA). Each method has its limitations: commercially available particle counters may underestimate EVs smaller than their limit of detection (e.g., ~50 – ~100 nm depending on the instrument), while total protein quantification assumes a perfect correlation between protein and particle counts.

2. Although quantification using a standard curve is recommended, it is not strictly necessary provided that (1) all measurements are performed within the same experiment and (2) both Ex and CoEx readings fall within the assay's dynamic range (i.e., are neither saturated nor below the detection threshold).